\documentclass[aps,prd,preprint,nofootinbib]{revtex4-2}
\usepackage{graphicx}
\usepackage{amssymb}
\usepackage{amsmath}
\usepackage{subcaption}
\usepackage{multirow}
\usepackage{booktabs}
\begin{document}


\title{Impact on time delays due to milli-lensing by subhalos on lensed gravitational waves}


\author{Ishan Swamy}
\email{ishanswamy2507@gmail.com}

\thanks{Corresponding Author}
\affiliation{Department of Physics, Dr. Vishwanath Karad MIT World Peace University, Pune, India}

\author{Anupreeta More}
\email{anupreeta@iucaa.in}
\affiliation{Inter-University Centre for Astronomy and Astrophysics, Post Bag 4, Ganeshkhind, Pune, India}
\affiliation{Kavli IPMU (WPI), UTIAS, The University of Tokyo, Kashiwa, 277-8583, Chiba, Japan}

\author{Ojas Patil}
\affiliation{Department of Physics, Indian Institute of Science Education and Research, Dr. Homi Bhabha Road, Pune, India}


\date{\today}

\begin{abstract}
Dark matter substructure properties such as their mass function and spatial distribution depend on the nature of dark matter and are strong tests of the cosmological model. Like luminous matter, these dark matter substructures cause gravitational lensing affecting observables such as time delays. Given the millisecond-level timing precision achievable with current and future gravitational-wave (GW) detectors, gravitationally lensed GWs provide a powerful probe of dark matter substructure, especially, at the lower end of the mass function. In contrast, the optical (or electromagnetic) observations can provide a precision of a few hours and are thus, not sensitive to deflections from the less massive subhalos. In this work, we investigate the impact of realistic population of DM subhalos ($10^6-10^9 M_\odot$) on the lensing time delays for two of the typical strong lens configurations, a fold and a cusp, seen in quadruply lensed sources for a galaxy-scale lens. We find that the subhalos with NFW density profiles cause perturbations of the order of few hours to the lensing time delays between the macro-lensed images produced by the main lensing galaxy. These time delay "anomalies", if not accounted for, may affect the results of strong lens searches conducted on the GW data by the LIGO--Virgo--Kagra collaboration. Lastly, for the 200 realisations analysed per fold and cusp lens systems, we find that the NFW subhalos produced no additional images of their own called milli-images. Statistical studies are needed to better determine the expected impact on the lensed GW time delays and (non-)detection of milli-images.
\end{abstract}


\maketitle

\section{Introduction}
Decades of observational studies have established the $\Lambda$CDM model as the current best-fit cosmology \citep [e.g.,][]{Dodelson03}. A combination of gravitational instability in cold dark matter (CDM) and quantum fluctuations during inflation led to a hierarchy of structure formation in this model. The success of this model lies in its accurate predictions backed by observations of cosmic microwave background (CMB) anisotropies \citep[e.g.,][]{Planck15, planck20a, planck20b}, and galaxy formation \citep[e.g.,][]{Anderson14, Vogel14, Schaye14, DESI25} . On smaller scales, however, there is a discrepancy between CDM simulation predictions and observations of galactic structure, particularly the missing satellite halo problem \citep[e.g.,][]{Kauffmann93, Moore99, Klypin99, Kravtsov10}. Simulations predict a large number of small mass subhalos bound within parent halos of galaxies, with their masses ranging from $10^3 - 10^{10}$~M$_\odot$ \citep[e.g.,][]{Diemand08, Stadel09, Navarro10}, but observations suggest a number less by a factor of 10. This problem has led to numerous suggested solutions, such as modifying the nature of dark matter. This involves models such as the warm dark matter\citep[e.g.,][]{Bode01, Abazajian06}, self-interacting dark matter \citep[e.g.,][]{Spergel00}, and fluid dark matter \citep[e.g.,][]{Khoury15}. Another class of solutions inhibits star formation, resulting in a purely dark substructure population \citep[e.g.,][]{Klypin99, Bullock01}.

One of the most prominent observational methods to detect dark matter is using gravitational lensing \citep[e.g.,][]{Mao98, Metcalf01, Chiba02, Dalal02, Moustakas03, Keeton03, Zentner03, Zentner05, Schneider06, More09, Vegetti10, Vegetti12}, because it is sensitive to both dark and luminous matter (see \cite{Liao22} for a review). Strong gravitational lensing occurs when electromagnetic (EM) waves encounter a massive lens (such as a galaxy or cluster) along its line-of-sight, resulting in observable phenomena such as magnification, phase shifts, and the production of multiple signals (images) arriving at different times. Small substructure affects these observations by changing the positions and/or flux ratios of images, with the fraction of projected mass required to satisfy the anomalous flux-densities of the lensed images being $0.0006-0.07$ of the parent halo at 90\% confidence level \citep[e.g.,][]{Dalal02}. Recent analysis of strongly lensed quasars have constrained the projected substructure mass fraction further to $~0.018-0.056$ near lensed images \citep{Hsueh19, Gilman19a}. However, for the inner part of halos, within a few kpc from the centre, the expected substructure fraction is only $\lesssim 0.3\%$ \citep[e.g.,][]{Mao04, Diemand07}. Further, ALMA observations have provided evidence of a $\approx 10^9 ~\rm{M}_\odot$ subhalo near one of the images, showing that interferometric measurements of strong gravitational lenses can be used to asses substructure induced lensing effects \citep{Hezaveh16}. The previous decade has also seen multiple studies attempting to constrain the substructure properties such as their mass models and density profiles \citep[e.g.,][]{Xu_15, Han16, Brehmer19, Gilman19, Gilman19b, Minor21}.

Beyond flux ratios of images, time delays between macro-images have proven to be an effective method to infer substructure properties \citep[e.g.,][]{Keeton09, Abe25, Gannon25}. Time delays, which depend directly on the lens potential, have been used to measure cosmological parameters too such as the Hubble constant ($H_0$), in an attempt to explain the discrepancy between the two $H_0$ values \citep[e.g.,][]{Chen19, Birrer19, Wong19, Shajib19, Gilman20}.

Alongside gravitational lensing, Einstein's general relativity also predicts that perturbations of spacetime can create waves called gravitational waves (GWs, see \cite{Sathya09} for a review). GWs from sources such as merging binary black holes, also get lensed by intervening objects, similar to lensed EM sources \citep[e.g.,][]{Rojo}. While current detectors (LIGO/Virgo/KAGRA) are limited to the relatively local universe ($z \lesssim 0.1$ for typical binaries), the next generation of ground-based observatories will fundamentally transform the landscape. The Einstein Telescope \citep{Regimbau12} and the Cosmic Explorer \citep{Abbott17} are projected to detect hundreds of lensed GW signals per year \citep[e.g.,][]{Pirkowska13, Ding15, Ng18, Li18, Wong19, Yang21}. These detectors will extend our horizon to high redshifts ($z>2$), effectively mapping the lensed GW population across cosmic time.

It is predicted that strongly lensed GW events produce image pairs with delays of less than a day, while demagnified signals can lag by up to 100 days  \cite{Oguri18}. Crucially, lensed GWs can provide a much more precise measurement of these time delays, with a time resolution of the order of a millisecond, in comparison to an EM signal, which has a precision of less than a day \citep[e.g.,][]{Liao17, Birrer25}.  
However, since GW detectors have broad sky localization, the identification of an EM counterpart remains critical to pinpointing the lens, enabling researchers to constrain dark matter substructure properties with a level of accuracy currently unattainable by light alone. Motivated by these developments, we investigate the impact of dark matter subhalos on the time delays of strongly lensed images in  galaxy-scale lens systems that mimick two real lenses, namely, PG 1115+080 and RX J1131-1231. 

The remainder of this paper is organized as follows. Section~\ref{sec:lenfw} outlines our lensing framework summarizing how we setup and simulate our lens systems. 
In Section~\ref{sec:subfw}, we describe the different models, choices and assumptions to characterise the subhalo population included in our framework. Section~\ref{sec:lmimp} details the macro models for two mock lens systems mimicking PG1115+080 and RX J1131-1231, as well as the two specific subhalo populations. 
We present our findings in Section~\ref{sec:results}, with a particular focus on how each subhalo population model influences time delays and other lensed observables such as magnifications. Finally, we provide a summary of our results and concluding remarks in Section~\ref{sec:conc}. 

Our calculations involve a combination of {\tt Astropy} \citep{astropy:2013, astropy:2018, astropy:2022},  {\tt Glafic}\footnote{https://github.com/oguri/glafic2} \citep{Oguri10}, {\tt Colossus}\footnote{https://bitbucket.org/bdiemer/colossus} \citep{Diemer18} and {\tt Lenstronomy}\footnote{https://github.com/lenstronomy/lenstronomy} \citep{Birrer15, Birrer18, Birrer21}. All quantities are reported in physical units unless otherwise specified. 


\section{Lensing Framework}
\label{sec:lenfw}

\begin{figure}
    \centering
    \includegraphics[scale =0.33]{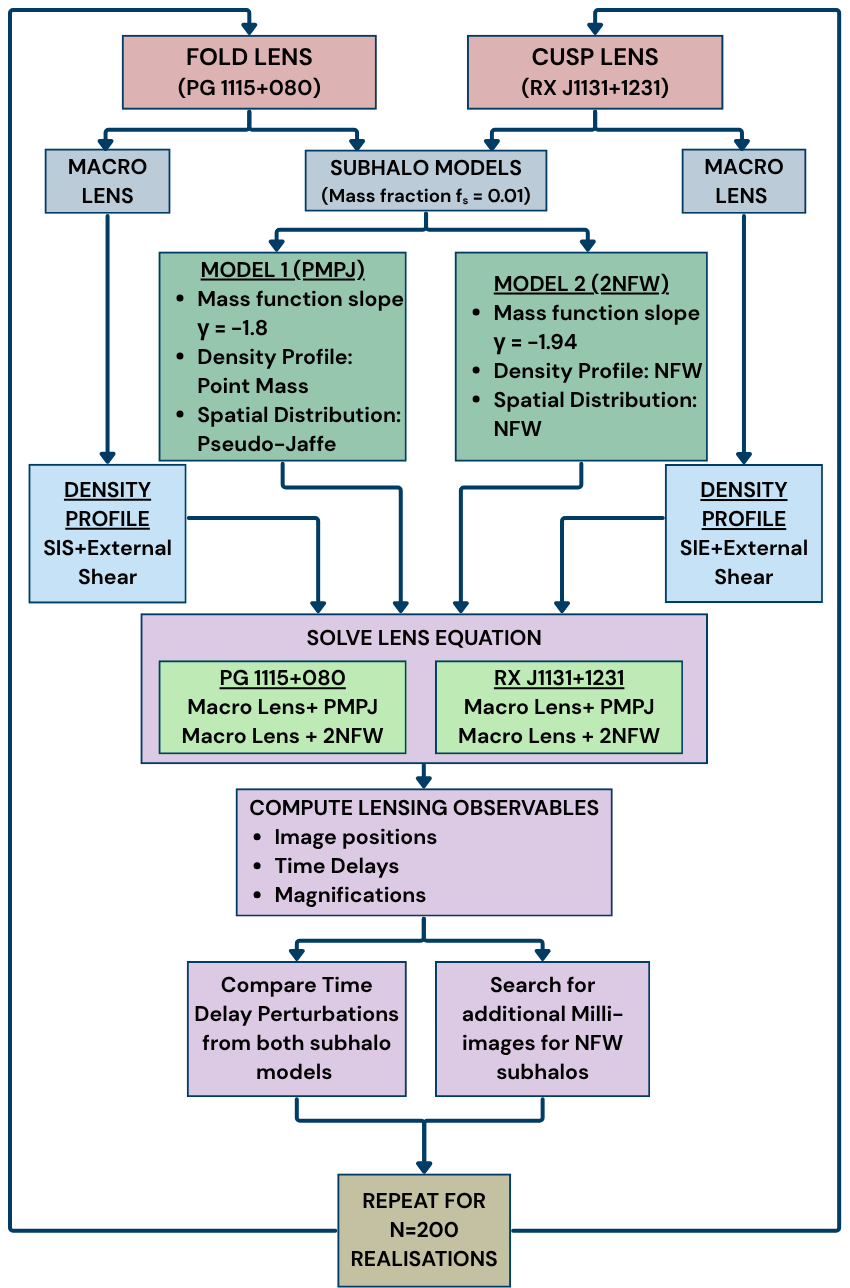}
    \caption{Flowchart illustrating the lensing framework.}
    \label{flwchrt}
\end{figure}

In this section, we give a brief summary of the overall framework (see also Fig.~\ref{flwchrt}) before diving into the details. We consider two mock lens systems with quadruple images, a fold and a cusp lens, representing typical lens configurations. 
For each mock lens, the lens mass model has two components to them -- a smooth macro-lens and a population of dark matter subhalos referred to as milli-lenses. We introduce subhalos by replacing a fraction of smooth halo mass $f_{\rm sub}$ with subhalos, and consider two distinct subhalo population models - 
a baseline model following \citet[][henceforth, KM09]{Keeton09} and a realistic model following the Navarro-Frenk-White \citep[NFW,][, see Section~\ref{submod} for details]{Navarro96}. 
Thus, we explore four distinct types of mock lenses in this work.

For each of the four mock lenses, we obtain the macro-lensing observables, namely the image positions, time delays, and magnifications by solving the lens equation. 
The lens equation is given by 
\begin{equation}
\boldsymbol\beta = \boldsymbol\theta - \boldsymbol\alpha(\boldsymbol\theta)
\label{lenseq}
\end{equation}
where $\boldsymbol\beta$ and $\boldsymbol\theta$ denote the angular positions of the source and the lensed images, respectively, and  $\boldsymbol\alpha(\boldsymbol\theta)$ is the scaled deflection angle. The scaled deflection angle can be obtained from the lensing potential using $\boldsymbol \alpha(\boldsymbol \theta)$=$\nabla\phi(\boldsymbol \theta)$. The lensing potential, which is a measure of how matter deflects light, is given as,  
\begin{equation}
\begin{aligned}
    \phi(\boldsymbol{\theta}) &= \frac{1}{\pi} \int \kappa(\boldsymbol{\theta}') \ln|\boldsymbol{\theta} - \boldsymbol{\theta}'| \, \mathrm{d}^2\theta' \\ 
\end{aligned}
\label{poteq}
\end{equation}
where $\boldsymbol \theta'$ denotes the angular position of an infinitesimal mass element on the lens plane, while $\boldsymbol \theta$ is the angular position at which the lensing potential is evaluated (usually the image position). The convergence $\kappa(\boldsymbol{\theta}') = \Sigma(\boldsymbol{\theta}')/ \Sigma_{\mathrm{cr}}$ is defined as a dimensionless surface density where the surface density, $\Sigma(\boldsymbol{\theta}')$, can be obtained by integrating the three dimensional density, $\rho(\boldsymbol \theta', z)$, and the critical surface density is given by $\Sigma_{\rm cr} = (c^2D_{\rm s})/(4\pi GD_{\rm l}D_{\rm ls})$. Here, the $D_{\rm l}$, $D_{\rm s}$, and $D_{\rm ls}$ are the angular distances between the observer and lens, the observer and source, and the lens and source, respectively.

As mentioned earlier, our lens model includes the smooth macro-lens and a population of subhalos representing the milli-lenses. Hence, the lensing potential is a sum of the contributions from the macro lens ($\phi_{\rm macro}$), parametrized by its total mass $M_{\rm tot}$, and the population of $N$ subhalos ($\phi_{\rm sub}$), given by
\begin{equation}
    \phi_{\rm tot} = \phi_{\rm macro}(\boldsymbol \theta; M_{\rm tot}) + \sum_{k=1}^{N}\phi_{\rm sub}(\boldsymbol \theta; \boldsymbol r_k, m_k)
    \label{lenspottot}
\end{equation}


Using Eqs.~\ref{lenseq}-\ref{lenspottot}, we can now determine the time delays ($\Delta\tau_{ij, \rm tot}$) between any two images $i$ and $j$ as
\begin{equation}
\Delta\tau_{ij, \rm tot} = t_0\left[\frac{1}{2}(|\boldsymbol{\theta}_i - \boldsymbol{\beta}|^2 - |\boldsymbol{\theta}_j - \boldsymbol{\beta}|^2) - \Delta\phi_{ij,\rm tot}\right]  
\label{tdeleq}
\end{equation}
The factor $t_0$ is given by
\begin{equation}
    t_0 = \frac{1+z_{\rm l}}{c} \frac{D_{\rm l}D_{\rm s}}{D_{\rm ls}}
\end{equation}
where $z_{\rm l}$ is the lens redshift. Our analysis focuses on the additional time delays arising due to the subhalos in the main lens system, which can be simply estimated as $t_0*\Delta\phi_{ij, \rm sub}$. 

In Eq.~\ref{poteq} and Eq.~\ref{lenspottot}, the lensing potential associated with an individual k-th subhalo is parametrized by its position ($\boldsymbol r_k$) on the lens plane and total mass ($m_k$), along with its internal density profile $\rho_{\rm sub}$.
As a result, to define the subhalo population completely, we need to specify the subhalo mass function (SHMF), the spatial distribution of the subhalos within the halo and the radial density profile of each subhalo.

The SHMF is assumed to follow a power-law distribution \citep[e.g.,][]{Ghigna00, Helmi02, Gao04, Diemand07},
\begin{equation}
    {\rm d}N/{\rm d}m \propto m^{\gamma}
    \label{massfunc}
\end{equation}
where $\gamma$ is the power-law slope. While, the total number of subhalos can be obtained by integrating Eq.~\ref{massfunc}, we instead adopt the simple approximation introduced by \citep{Keeton09}, where the mean subhalo count is expressed as, 
\begin{equation}
	\langle{N}\rangle = \frac{f_{\rm sub} M_{\rm tot}}{\langle{m}\rangle}
    \label{totnum}
\end{equation}
for a fraction $f_{\rm sub}$ of total halo mass ($M_{\rm tot}$) assigned to subhalos. Here, $\langle{m}\rangle$ is the mean subhalo mass \cite{Keeton09}, obtained by integrating Eq.~\ref{massfunc}, 
\begin{equation}
    \langle{m}\rangle = \frac{1+\gamma}{2+\gamma} \frac{m_2^{2+\gamma} - m_1^{2+\gamma}}{m_2^{1+\gamma} - m_1^{1+\gamma}}
    \label{meanmass}
\end{equation}
where $m_1$ and $m_2$ are the lower and upper mass limits of our subhalos. Further details of the specific model choices and assumptions for the subhalo population are described in Section~\ref{sec:lmimp}.


Lastly, we also investigate the formation of milli-images for the NFW subhalo model for each lens configuration. All of the above steps are repeated for 200 independent Monte Carlo realizations. 

\section{Framework for the Subhalo Population}
\label{sec:subfw}
We adopt a probabilistic approach to substructure modelling by defining three fundamental components: the mass function of subhalos, the spatial distribution of subhalos, and the density profile of each subhalo. These components collectively determine the gravitational potential and the resulting lensing perturbations of the subhalo population.


\subsection{Mass Function}
\label{shmf}
The SHMF, as discussed earlier (see Section~\ref{sec:lenfw}), can be used to obtain the individual mass of subhalos, given as
\begin{equation}
    {\rm d}N/{\rm d}m \propto m^{\gamma}
\end{equation}
where $N$ is the total number of subhalos of mass $m$, and $\gamma$ is the power-law slope. 
The corresponding normalised probability density \citep{Keeton09a, Keeton09} is
\begin{equation}
p(m)=\frac{(1+\gamma)m^\beta}{m_{\gamma}^{1+\gamma}-m_{\min}^{1+\gamma}},
\end{equation}
in the range $m_{\rm min} < m < m_{\rm max}$ for $\gamma \neq -1$. Individual subhalo masses are generated using inverse-transform sampling. Specifically, drawing a uniform random number $P_m \in [0,1]$ and equating it to the cumulative distribution,
\begin{equation}
P_m=\int_{m_{\min}}^{m}p(m'),dm',
\end{equation}
giving
\begin{equation}
P_m=\frac{m^{1+\gamma}-m_{\min}^{1+\gamma}}{m_{\max}^{1+\gamma}-m_{\min}^{1+\gamma}}
\label{mcdf}
\end{equation}

Inverting Eq.~\ref{mcdf}, each subhalo gets assigned a mass as,
\begin{equation}
	m = \left[P_m(m)\left(m_{\rm max}^{1+\gamma} - m_{\rm min}^{1+\gamma}\right) + m_{\rm min}^{1+\gamma}\right]^{1/(1+\gamma)}
    \label{massdist}
\end{equation}


\subsection{Spatial Distribution}
\label{sd}

The projected spatial distribution of subhalos influences the lensing cross-section and the extent to which the image positions, magnifications, and time delays are perturbed. Following KM09, we adopt the Pseudo-Jaffe (PJ) model as a reference. We also consider a more realistic NFW profile for the projected radial distribution of subhalos, motivated by cosmological simulations \citep{Gao04, Springel08}.


\subsubsection{Pseudo-Jaffe spatial distribution}
\label{pjsd}
For the Pseudo-Jaffe model, the radial position, $r$, of each subhalo, measured from the centre of the macro lens, is generated by sampling from the Pseudo-Jaffe radial distribution. The cumulative probability distribution for the radius is given as \citep{Keeton09}, 
\begin{equation}
	P_r(r) = 1 + \frac{r}{a} - \left(1+\frac{r^2}{a^2}\right)^{1/2} \,.
\end{equation}
Using inverse transform sampling, the radius is obtained from a uniformly distributed random variable $P_r \in [0,1]$
\begin{equation}
    r_k = a\frac{P_r(2-P_r)}{2(1-P_r)}\,.
    \label{radial}
\end{equation}
The sampled radial positions are combined with random azimuthal angles (from 0 to $2\pi$) assuming isotropy to generate the projected positions of subhalo population.


\subsubsection{NFW spatial distribution}
\label{nfwsd}
We generate the three-dimensional spatial distribution of subhalos according to the cumulative NFW profile given as,
\begin{equation}
    F_{\rm NFW}(x, x_{\max}) = \frac{\ln(1 + x) - \frac{x}{1 + x}}{\ln(1 + x_{\max}) - \frac{x_{\max}}{1 + x_{\max}}}
\label{scdf}
\end{equation}
where $x=r_k^{\rm 3D}/R_{\rm s}^{\rm host}$, $r_k^{\rm 3D}$ is a three-dimensional distance of the $k$-th subhalo from the centre of the host galaxy. The maximum dimensionless radius is defined as $x_{\rm max}=R_{200}^ {\rm host}/R_{\rm s}^{\rm host}$ where $R_{200}^{\rm host}$ is the radius at which the dark matter halo density is 200 times the critical density of the universe. To sample the radial position, a random number $P_r \in [0,1]$, drawn from a uniform distribution, is assigned to the cumulative probability,
\begin{equation}
P_r=F_{\rm NFW}(x,x_{\max}).
\end{equation}
Thus, $x$ is obtained by numerically solving
\begin{equation}
\ln(1+x)-\frac{x}{1+x} = P_r\left[\ln(1+x_{\max})-\frac{x_{\max}}{1+x_{\max}}\right]
\label{nfw_sampling}
\end{equation}
which gives,
\begin{equation}
r_k^{\rm 3D} = xR_{\rm s}^{\rm host}
\end{equation}
The corresponding three-dimensional position is then assigned an isotropic orientation and projected onto the lens plane to obtain $r_k$.


\subsection{Mass Density Profile}
\label{dp}
We consider two choices to define the density profile of each subhalo. The point mass model is used for simplicity and serves as a reference to the work of KM09 whereas the NFW profile is used as it is a more realistic density profile for the dark matter subhalos \citep{Navarro04, Springel08}.  


\subsubsection{Point Mass Density}
\label{pmdp}
In the point-mass approximation, the projected surface density of the $k$-th subhalo is represented by
\begin{equation}
    \Sigma(\boldsymbol \theta') = m \delta^{(2)}(\boldsymbol{\theta'} - \boldsymbol{r_k})
\end{equation}
where $\delta^{(2)}$ is the two-dimensional Dirac delta function.
Substituting this expression into Eq.~\ref{poteq} results into the lensing potential as,
\begin{equation}
    \phi_{\rm sh} = \frac{\hat{m}}{\pi} \ln r
\end{equation} 
where $r = |\boldsymbol{\theta} - \boldsymbol{r_k}|$ is the angular separation between the evaluation point $\theta$ and the centre of the subhalo. Here the scaled mass $\hat{m}$ for a subhalo of mass $m_k$ and Einstein radius $R_{{\rm Ein},k}$ is \citep{Keeton09}
\begin{equation}
    \hat{m_k} \equiv \frac{m_k}{\Sigma_{\rm crit}} = \pi R^2_{{\rm Ein},k}\,.
    \label{scaledmass}
\end{equation}
For a realisation containing N point-mass subhalos, the total perturbation to the lensing potential is obtained by summing the contribution from each subhalo. Consequently, the perturbation to the potential difference between two macro-images located at $\boldsymbol \theta_i$ and $\boldsymbol \theta_j$ is
\begin{equation}
\begin{aligned}
     \Delta\phi_{ij, \rm sub} &= \sum_{k=1}^{N} \phi(\boldsymbol{r}_k,m_k), \\
     \textnormal{where} \quad \phi(\boldsymbol{r_k},m_k) &= \frac{\hat{m_k}}{\pi} \ln \frac{|\boldsymbol{\theta}_i - \boldsymbol{r_k}|}{|\boldsymbol{\theta}_j - \boldsymbol{r_k}|}
\end{aligned}
\label{pointpot}
\end{equation}
The total potential perturbation is, therefore, determined by the sum of the individual contributions of all subhalos in a given realization.


\subsubsection{NFW Density}
\label{nfwdp}
The NFW density profile for a subhalo is described as, 
\begin{equation}
	\rho(r_{\rm int}^{\rm sub}) = \frac{\rho_s^{\rm sub}}{(r_{\rm int}^{\rm sub}/R_s^{\rm sub})(1 + r_{\rm int}^{\rm sub}/R_s^{\rm sub})^2}
    \label{eq:nfwprof}
\end{equation}

where $r_{\rm int}^{\rm sub}$ is the radial distance from the centre of the subhalo (or the internal radius), and $\rho_s^{\rm sub}$ is the characteristic density. The characteristic density $\rho_s^{\rm sub}$ can be expressed as
\begin{equation}
	\rho_s^{\rm sub}=    \frac{200}{3}\rho_{\rm crit}\frac{(c_{200}^{\rm sub})^{3}}{\ln(1+c_{200}^{\rm sub})-\frac{c_{200}^{\rm sub}}{1+c_{200}^{\rm sub}}}.
\end{equation}
where the concentration $c_{200}=R_{200}^{\rm sub}/R_{\rm s}^{\rm sub}$ and $\rho_{\rm crit}$ is the critical density of the universe.

The deflection angle is determined by,
\begin{equation}
\alpha_{\rm sub}(x_{\rm sub})
=
\frac{4\kappa_s^{\rm sub}R_{s,\rm ang}^{\rm sub}}{x_{\rm sub}}
\left[
\ln\left(\frac{x_{\rm sub}}{2}\right)
+
g(x_{\rm sub})
\right]
\label{nfw_deflection}
\end{equation}
and
\begin{equation}
g(x_{\rm sub}) =
\begin{cases}
\displaystyle
\frac{1}{\sqrt{x_{\rm sub}^2-1}}
\arctan\sqrt{x_{\rm sub}^2-1},
& x_{\rm sub}>1, \\[10pt]
1,
& x_{\rm sub}=1, \\[10pt]
\displaystyle
\frac{1}{\sqrt{1-x_{\rm sub}^2}}
\operatorname{arctanh}\sqrt{1-x_{\rm sub}^2},
& x_{\rm sub}<1.
\end{cases}
\label{nfw_f}
\end{equation}
here $x_{\rm sub} = r_{\rm int,ang}^{\rm sub}/R_{\rm s,ang}^{\rm sub}$ and $R_{\rm s,ang}^{\rm sub}= R_s^{\rm sub}/D_{\rm l}$ is the angular scale radius.

The lensing potential is therefore,
\begin{equation}
    \phi_{\rm sub}(x_{\rm sub})
    =
    2\kappa_{\rm s}
    \left(R_{s,\rm ang}^{\rm sub}\right)^2
    h(x_{\rm sub}),
\end{equation}
where
\begin{equation}
    h(x_{\rm sub}) =
    \begin{cases}
        \displaystyle
        \ln^2\left(\frac{x_{\rm sub}}{2}\right)
        -
        \operatorname{arccosh}^2\left(\frac{1}{x_{\rm sub}}\right),
        & x_{\rm sub}<1,
        \\[2ex]
        \displaystyle
        \ln^2\left(\frac{x_{\rm sub}}{2}\right)
        +
        \arccos^2\left(\frac{1}{x_{\rm sub}}\right),
        & x_{\rm sub}\geq 1.
    \end{cases}
\end{equation}

\section{Lens Model Implementation}
\label{sec:lmimp}
Here we give details of the implementation of  the models to generate the mock lens systems.
\subsection{Macro-Lens Models}

For the macro-lens of the fold lens configuration, we consider the model of PG 1115+080 \citep{Weymann80, Tonry98} which is fit with a Singular Isothermal Sphere (SIS) and an external shear. For the macro-lens of the cusp lens configuration, we consider the model of RX J1131-1231 \citep{Sluse03, Sluse07} which is fit with a Singular Isothermal Ellipsoid (SIE) and an external shear. The source for PG1115+080 is set at ($-$0".032, 0".118), while for RX J1131-1231 it is set at ($-$0".549, $-$0".142). The Table~\ref{lensprop} lists the values of the parameters of the macro-lenses and the Table~\ref{imageprop} gives properties of the four macro-images taken from KM09.

\begin{table}
	\centering
	\caption{Macro-Lens Properties. Position Angles are measured East of North.}
	\label{lensprop}
	\begin{tabular}{lcr} 
	\hline
	Properties & PG 1115+080 & RX J1131-1231\\
	\hline
	Einstein Radius (") & 1.16 & 1.85\\
	$z_{\rm l}$ & 0.31 & 0.295 \\
	$z_{\rm s}$ & 1.72 & 0.658 \\
      	Ellipticity & 0 & 0.16\\
        Position Angle & 0 & $-56^\circ$\\
        Shear & 0.12 & 0.12\\
        Shear Position Angle & $65^\circ$ & $-83^\circ$\\
        \hline
    \end{tabular}
\end{table}

\begin{table}
\centering
\caption{Image Properties of the mock lenses. The naming of each image corresponds to its arrival time, with 1 being the earliest and 4 being the latest.}
\label{imageprop}
    	\begin{tabular}{lcccr} 
	\hline
	Lens & Image No. &  x  &  y  & $\Delta \tau_{ij}$\\
	     & 		 & (") & (") & (days) \\
	\hline
	&1 & 0.343 & 1.360 & 0\\
	PG 1115+080&2 & $-$0.948 & 0.697 & 10.77\\
	&3 & $-$1.098 & $-$0.206 & 10.93\\
        &4 & 0.700 & $-$0.652 & 17.82\\
        \hline
        &1 & $-$1.717 & $-$1.697 & 0\\
	RX J1131-1231&2 & $-$2.334 & 0.612 & 0.25\\
	&3 & $-$2.305 & $-$0.577 & 1.22\\
        &4 &  0.796 & 0.315 & 120.08\\
        \hline
	\end{tabular}
\end{table}


\subsection{Milli-lens Models}
\label{submod}
As discussed above, the mass assigned to each subhalo, its spatial distribution, and density profile are the properties that differentiate the two models in our framework. We work with two subhalo models, namely the Point Mass - Pseudo Jaffe (PMPJ) model and the NFW - NFW (2NFW) model in the mass range $10^6 - 10^9 M_\odot$.



\subsubsection{Model PMPJ}

This model pertains to subhalos with a point mass density profile (Section~\ref{pmdp}), and with the SHMF (Eq.~\ref{massfunc}) slope set to $\gamma = -1.8$ and $f_{\rm sub} = 0.01$ following KM09. The total subhalo count in this model is calculated from Eq~\ref{totnum} to be 10030 for PG 1115+080 and 23399 for RX J1131-1231.

\begin{figure}
    \centering
    \includegraphics[scale = 0.08]{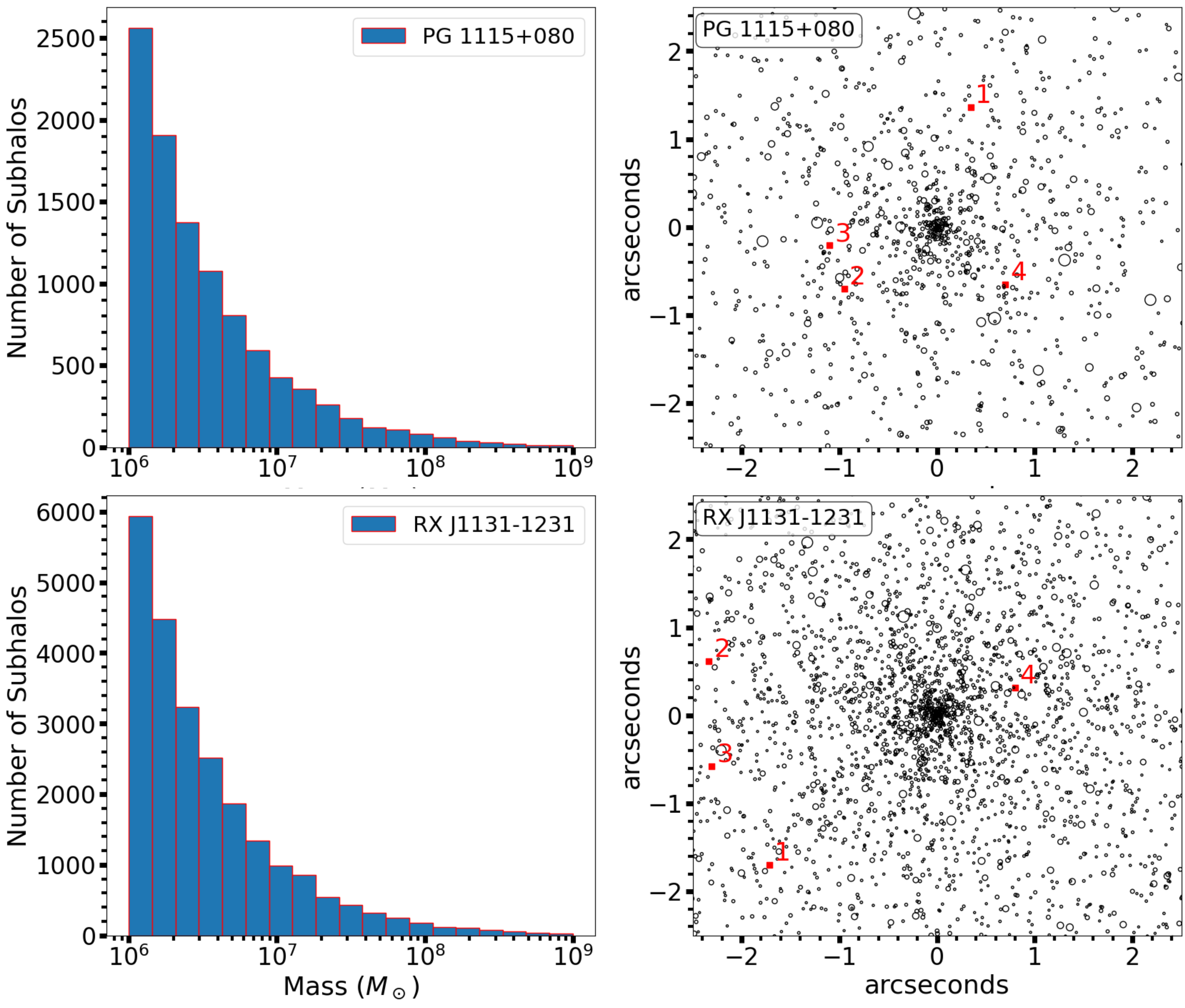}
    \caption{Number of subhalos as a function of their masses for PMPJ model (left) along with their spatial distribution (right) shown for each lens row-wise. The size of each point is scaled proportionally to the Einstein radius of the subhalo.}
    \label{PMPJdistfig}
\end{figure}

We assign a mass to each subhalo using Eq.~\ref{massdist}, and show the resulting distribution of the masses for the full population in Fig.~\ref{PMPJdistfig} (see left panel). 
It is evident that the number of lower-mass subhalos is exceptionally high.

The subhalos are then spatially distributed based on the Pseudo-Jaffe distribution (Section~\ref{pjsd}), with the scale radius $a$, taken as 300~kpc, also acting as a truncation radius. Each subhalo is assigned a position $(x,y)$ in the lens plane using Eq.~\ref{radial}. The right panels in Fig.~\ref{PMPJdistfig} show the subhalo distribution in the limits of $-2.5'' < x,y < 2.5'' $. The size of circular markers of the subhalos relates to their Einstein Radius calculated using Eq.~\ref{scaledmass}. In addition, the corresponding images formed due to the main lens have also been labelled in the right panel of Fig.~\ref{PMPJdistfig} based on their coordinates and time delays mentioned in Table~\ref{imageprop}.

\begin{figure}
\centering
	\includegraphics[scale = 0.1]{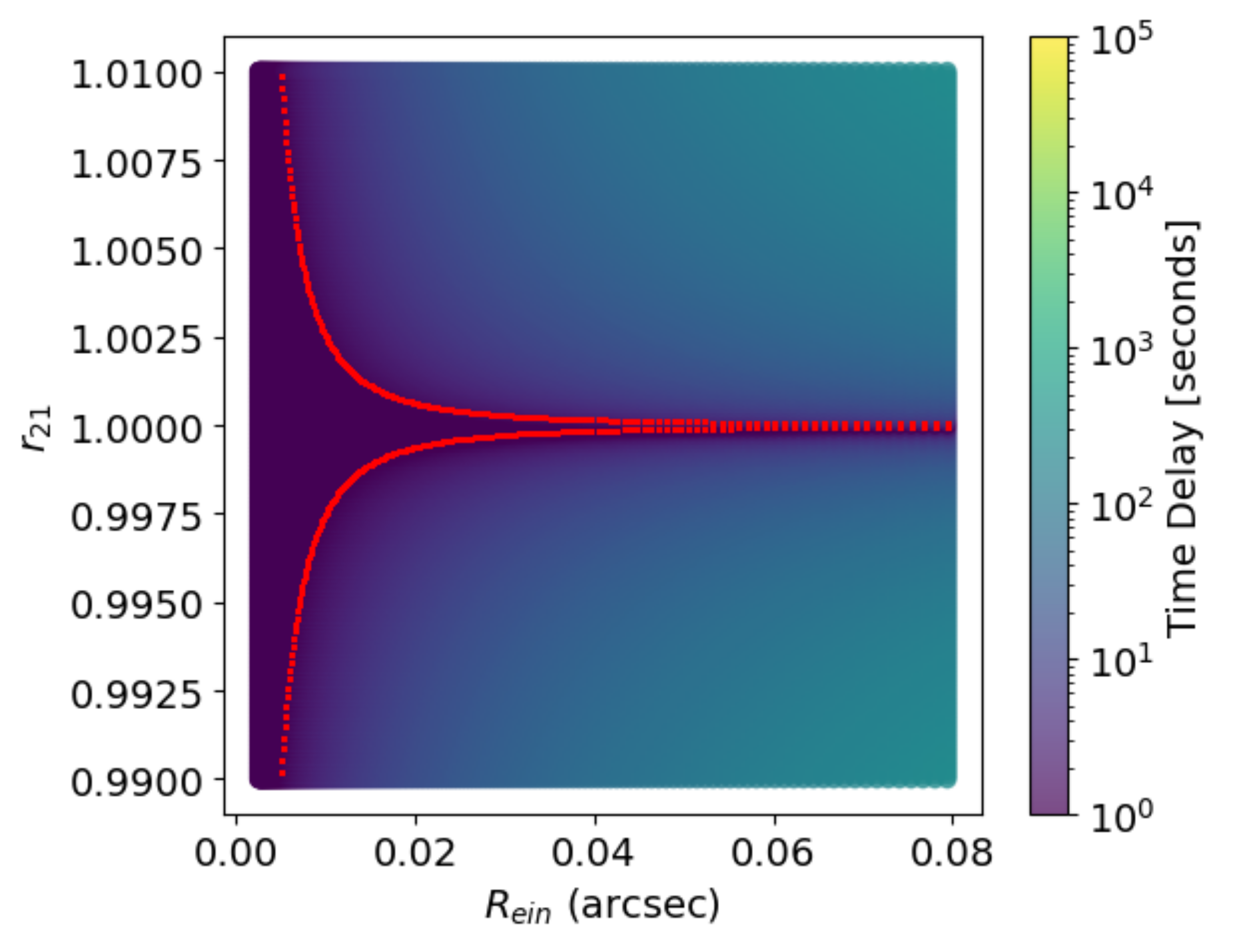}
    \caption{Time delay as a function of the Einstein radius and angular separation ($r_{21}$) between image pair 21. The parameter space inside the red boundary constitutes of subhalos that cause less than 100ms of perturbation, which will not be considered in lensing calculations.}
    \label{tdcolor}
\end{figure}

\begin{figure}
    \centering
    \includegraphics[scale = 0.5]{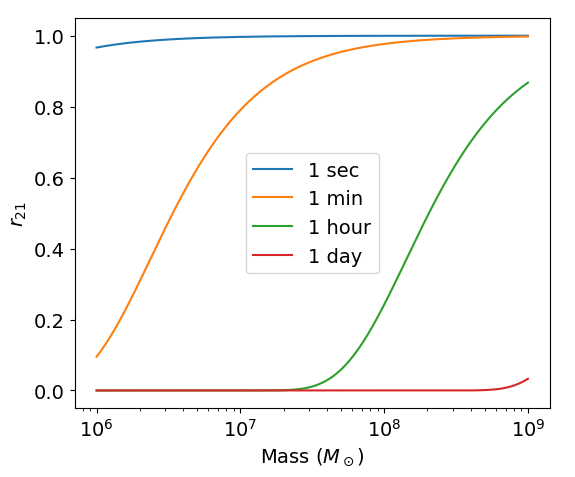}
    \caption{Variation of $r_{21}$ with mass for different time delays. The plot highlights the expected mass and distance from the images of a single subhalo, for a given time delay perturbation of 1 second, 1 minute, 1 hour and 1 day.}
\label{mass_vs_r21}
\end{figure}

\begin{figure*}
\centering
    \includegraphics[scale = 0.23]{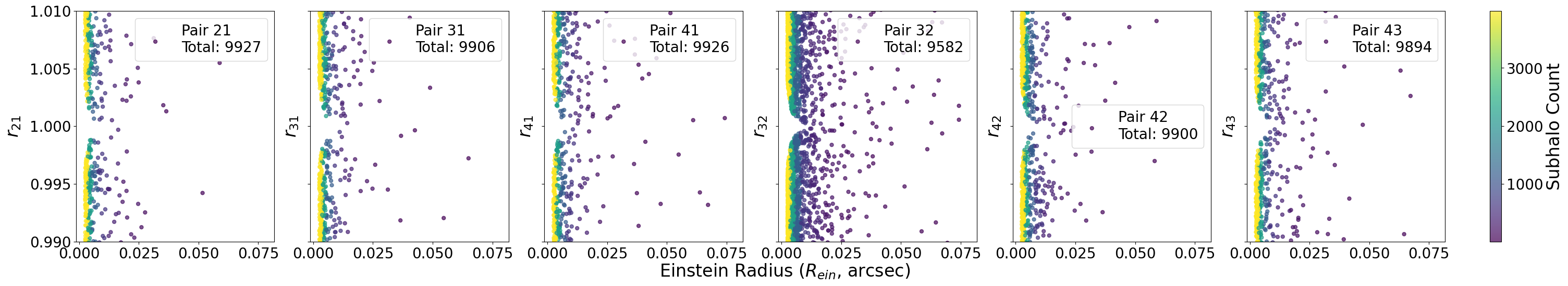}
    \caption{Einstein radius vs distance ratio for the subhalo population of for PG 1115+080 post filtering. The colour bar represents the subhalo counts.}
    \label{subcolor}
\end{figure*}

Before we solve the lensing observables for this model, we define a selection criterion to filter out subhalos whose time-delay perturbations are less than 100~ms. This is done in order to make the simulations computationally efficient, without losing much precision. Fig.~\ref{tdcolor} outlines, in red, the subhalos that contribute to less than 100~ms of time delay (purple region). Fig.~\ref{mass_vs_r21} illustrates the mass alongside the respective distance ratio of the image pair, which results in time delays of 1~second, 1~minute, 1~hour, and 1~day. This provides a broad understanding of the potential contribution from an individual subhalo with a specific mass located at a particular position. It can be noted that nearly all subhalos lead to a time delay of 1~second, whereas only a small region of higher mass subhalos causes a delay of 1~day. Although the analysis shown here is for the fold lens PG 1115+080, similar results are obtained for the cusp lens RX J1131-1231. Fig.~\ref{subcolor} confirms that the subhalos are filtered in both lenses, as the subhalos inside the red boundary of Fig.~\ref{tdcolor} are absent, with the subhalo count reflecting this as well.

\subsubsection{Model 2NFW}
Each subhalo in this model follows the NFW density profile (Eq.~\ref{eq:nfwprof}), which is less steep than the point mass and consequently, has a different SHMF slope. We test two SHMF slopes \citep{Gannon25}, the first being an unevolved $\gamma_{\rm u}=-1.93$, and set $f_{\rm sub,u}=0.0282$. The second is taken to be an evolved $\gamma_{\rm e}=-1.94$, and set $f_{\rm sub,e}=0.01$. Here, the unevolved case refers to the subhalos at infall mass, prior to being affected by processes such as tidal stripping and dynamical friction, whereas the evolved case measures their surviving masses at observation time.

\begin{figure}
    \centering
    \includegraphics[scale = 0.3]{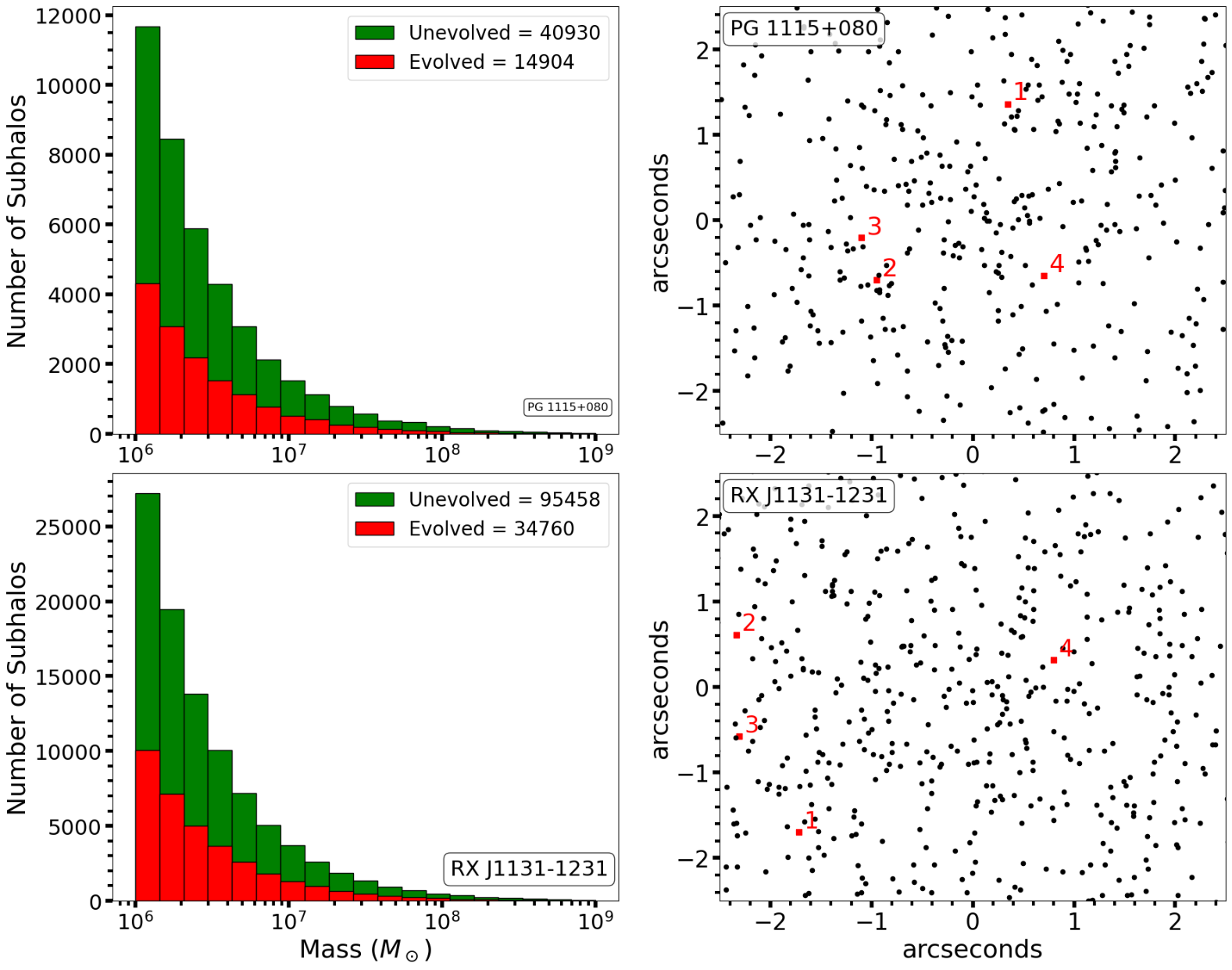}
    \caption{Number of subhalos per mass bin for unevolved and evolved SHMFs (left). Spatial distribution of subhalos for the evolved SHMF (right).}
    \label{nfwdistfig}
\end{figure}

The left panel of Fig.~\ref{nfwdistfig} shows the comparison of the subhalo masses and their total numbers for the two SHMFs (Eq.~\ref{totnum} and Eq.~\ref{massdist}). The unevolved mass function (green) leads to more than double the subhalos in the evolved SHMF, due to the higher mass fraction. Hereafter, we will be using the evolved SHMF for our second model for realistic predictions and 2NFW subhalo will refer to the evolved case.

The spatial distribution also corresponds to a NFW radially symmetric profile \citep[e.g.,][]{Navarro97, Keeton01}, given in Section~\ref{nfwsd}. The concentration $c_{200}^{\rm host}$ is calculated using {\tt Colossus} \citep{Diemer18} which follows \citet{Ishiyama21} by setting $R_{200}^{\rm host}$ as the virial radius of the host halo. Similar to Fig~\ref{PMPJdistfig}, the right panels of Fig~\ref{nfwdistfig} show the spatial distribution for the 2NFW model for both lenses.

\section{Results and Discussions}
\label{sec:results}
Here, we discuss the effects of the different subhalo population models on image properties such as time delays and magnifications for the two lens systems. We also investigate formation of milli-images due to the 2NFW subhalo model.

\subsection{Time Delay Perturbations}
We compute the time delay perturbations for a fold and a cusp type of lens and for each of these lenses two distinct subhalo population models are assumed. Our results for all four mock lenses are given in Table~\ref{tabresults}. The time delay perturbations are shown for all image pair combinations of a quad alongside the mean number count and mean masses of the subhalos. It is interesting that the mean perturbation due to PMPJ subhalos can be as high as 17 hours (Pair 41) in the PG 1115+080 lens and about 120 hours (Pair 41) in the RX J1131-1231 lens, highlighting the massive lensing effect of the subhalos. The 2NFW subhalos, on the other hand, cause a time delay of few hours in PG 1115+080 lens and a delay of less than a day for RX J1131-1231 on average. 

\begin{table*}
\centering
\caption{Results over 200 realisations}
\label{tabresults}
\begin{tabular}{lccccccccc}
\toprule

Lens & Image Pair 
& \multicolumn{2}{c}{Mean Subhalo} 
& \multicolumn{2}{c}{Mean Total Subhalo} 
& Macro-Only Time Delay 
& \multicolumn{2}{c}{Perturbation} \\

& 
& \multicolumn{2}{c}{Count} 
& \multicolumn{2}{c}{Mass ($10^{11}~M_\odot$)} 
& $\Delta\tau_{ij,\rm h}$ (days)
& \multicolumn{2}{c}{$\Delta\tau_{ij,\rm sh}$(hours)} \\
\midrule
& 
& PMPJ & 2NFW 
& PMPJ & 2NFW
&
& PMPJ & 2NFW \\

\midrule

\multirow{6}{*}{PG 1115+080}
& 21 & 9930 & 14904 & 1.20 & 1.20 & 10.712 & 7.38 & 1.70 \\
& 31 & 9919 & " & 1.20 & " & 10.859 & 9.47 & 2.31 \\
& 41 & 9914 & " & 1.20 & " & 17.713 & 14.59 & 2.97 \\
& 32 & 9552 & " & 1.19 & " & 0.147 & 2.09 & 0.60 \\
& 42 & 9886 & " & 1.20 & " & 7.001 & 7.21 & 1.26 \\
& 43 & 9901 & " & 1.20 & " & 6.854 & 5.12 & 0.66 \\

\midrule

\multirow{6}{*}{RX J1131-1231}
& 21 & 23141 & 34760 & 2.79 & 2.80 & 0.227 & 1.28 & 0.69\\
& 31 & 22918 & " & 2.78 & " & 1.177 & 3.52 & 0.82 \\
& 41 & 23219 & " & 2.79 & " & 119.422 & 120.02 & 21.53 \\
& 32 & 22891 & " & 2.78 & " & 0.95 & 2.24 & 0.12 \\
& 42 & 23214 & " & 2.79 & " & 119.195 & 118.74 & 20.83 \\
& 43 & 23220 & " & 2.79 & " & 118.245 & 116.49 & 20.71 \\

\bottomrule
\end{tabular}
\end{table*}
        
\begin{figure*}
\centering
\includegraphics[scale = 0.25]{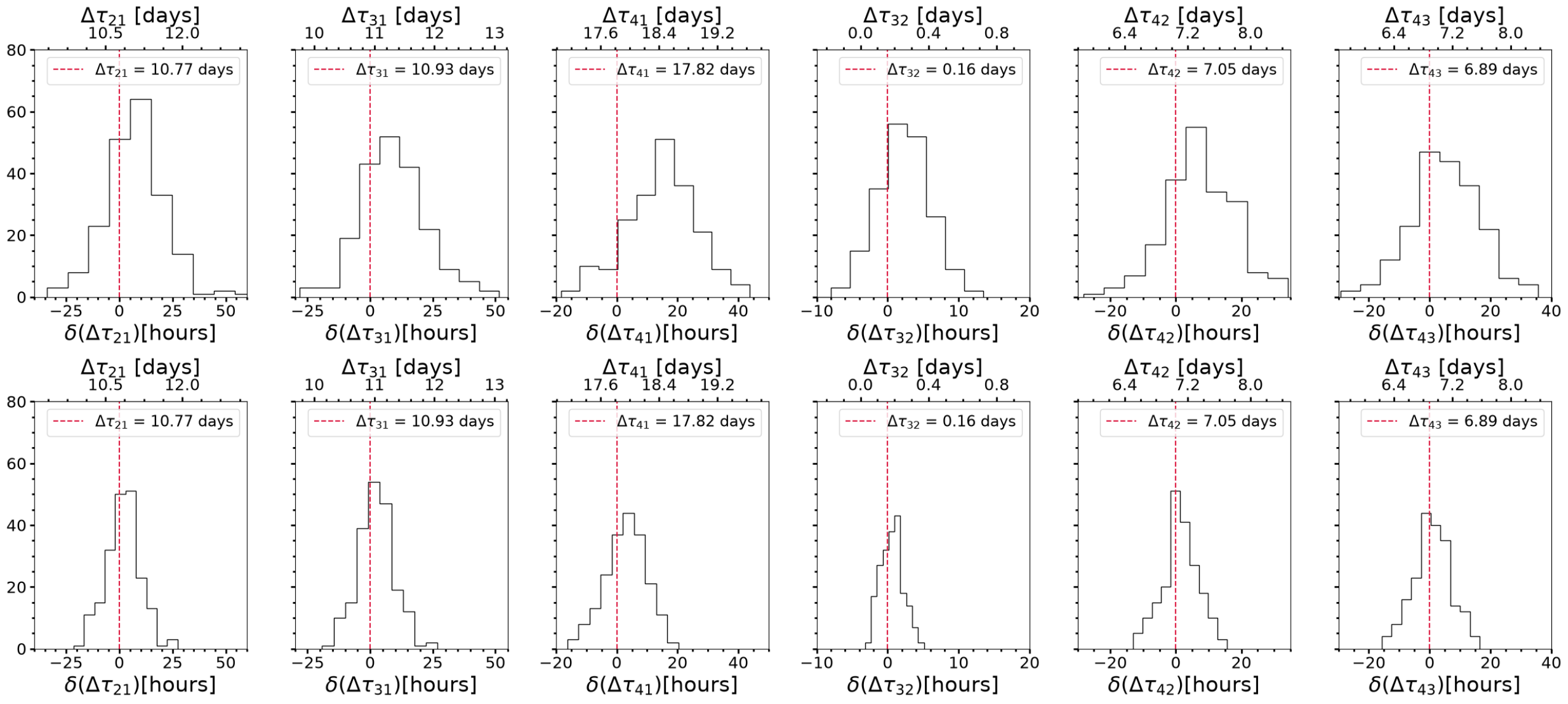}
\caption{Time delay perturbations in PG 1115+080 due to PMPJ (top) and 2NFW  (bottom) subhalos. In each panel, the bottom x-axis corresponds to the time delay due to the subhalos only, whereas the top x-axis corresponds to the total time delay due to the halo and subhalos. The red vertical line represents the total time delay due to the halo only without any subhalos.}
\label{pertpg}
\end{figure*}

\begin{figure*}
\centering
\includegraphics[scale = 0.27]{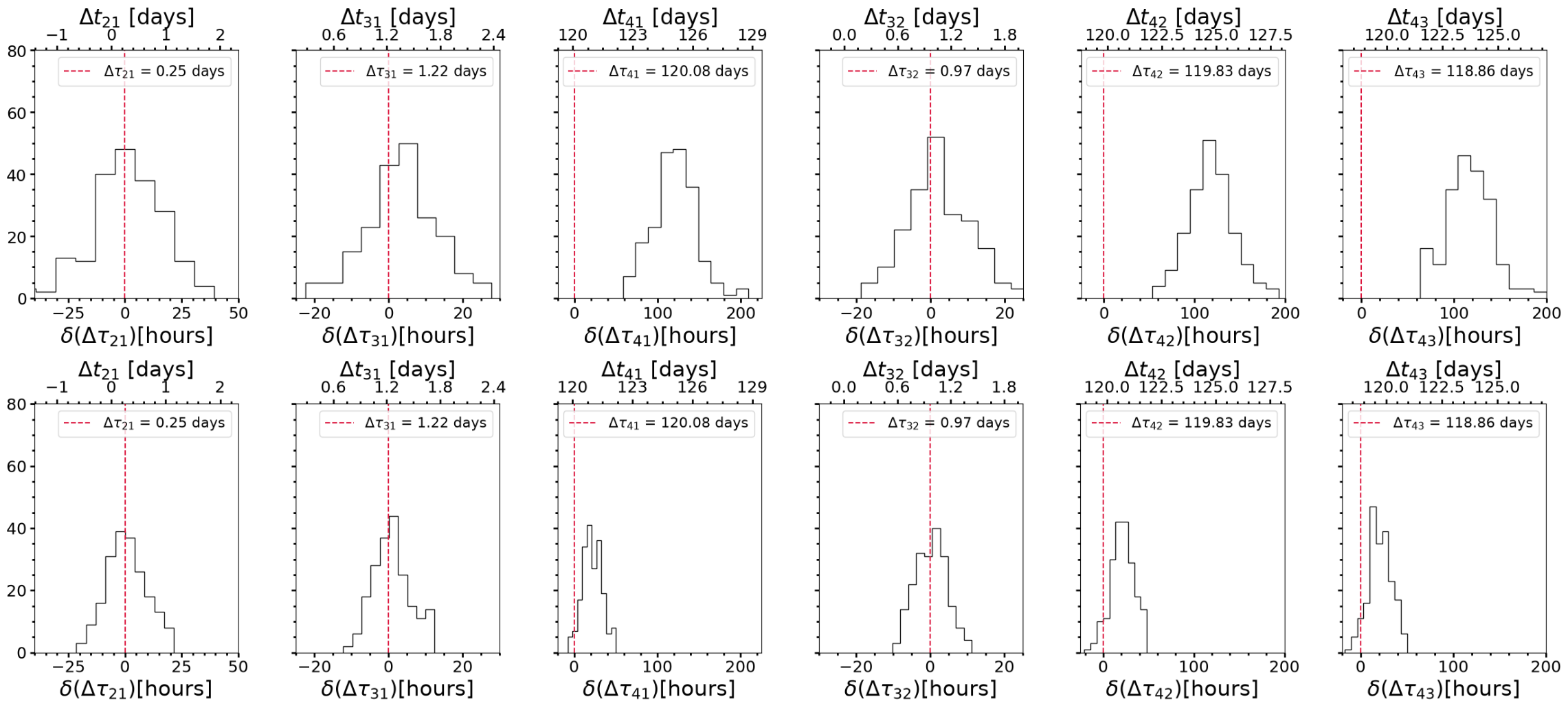}
\caption{Time delay perturbations in RX J1131+080 due to PMPJ (top) and 2NFW  (bottom) subhalos.}  
\label{pertrxj}
\end{figure*}

In Fig~\ref{pertpg}, it can be noted that some realisations can produce a perturbation of about 40~hours in system PG 1115+080 due to PMPJ (top panel) and about 20~hours due to 2NFW subhalos (bottom panel). A similar behaviour is visible in Fig~\ref{pertrxj} in system RX J1131-1231, where the maximum can be around 200~hours due to PMPJ (top panel), and about 50~hours due to 2NFW subhalos (bottom panel). We emphasise that, despite the differences in perturbation values, both model produce qualitatively similar results, with the maximum perturbations observed in pairs 21, 31 and 41 of PG 1115+080 (fold lens), and in pairs 41, 42 and 43 of RX J1131-1231 (cusp lens). A major difference, however, arises in pairs 41, 42, and 43, where in a few realisations, 2NFW  subhalos can lead to negative perturbations (indicating the latter image arriving earlier than the former image, which never occur in the PMPJ model. 

Quantitatively, it is observed that PMPJ subhalos produce higher perturbations consistently and consequently, larger time delays. The reason for 2NFW subhalos having shorter time delays compared to PMPJ subhalos is their shallower density profiles which corresponds to lower surface mass density (i.e. convergence). 
The other reason for the shorter time delay perturbation also lies in the spatial distribution of subhalos discussed in Appendix~\ref{appendix}. The significant outcome to note is that even a shallow and sparse 2NFW model is able to produce $\mathcal{O}$~(hours) perturbations, highlighting the need for subhalos to be accounted for during the modelling of lens systems.

\begin{figure*}
    \centering
    \includegraphics[scale = 0.23]{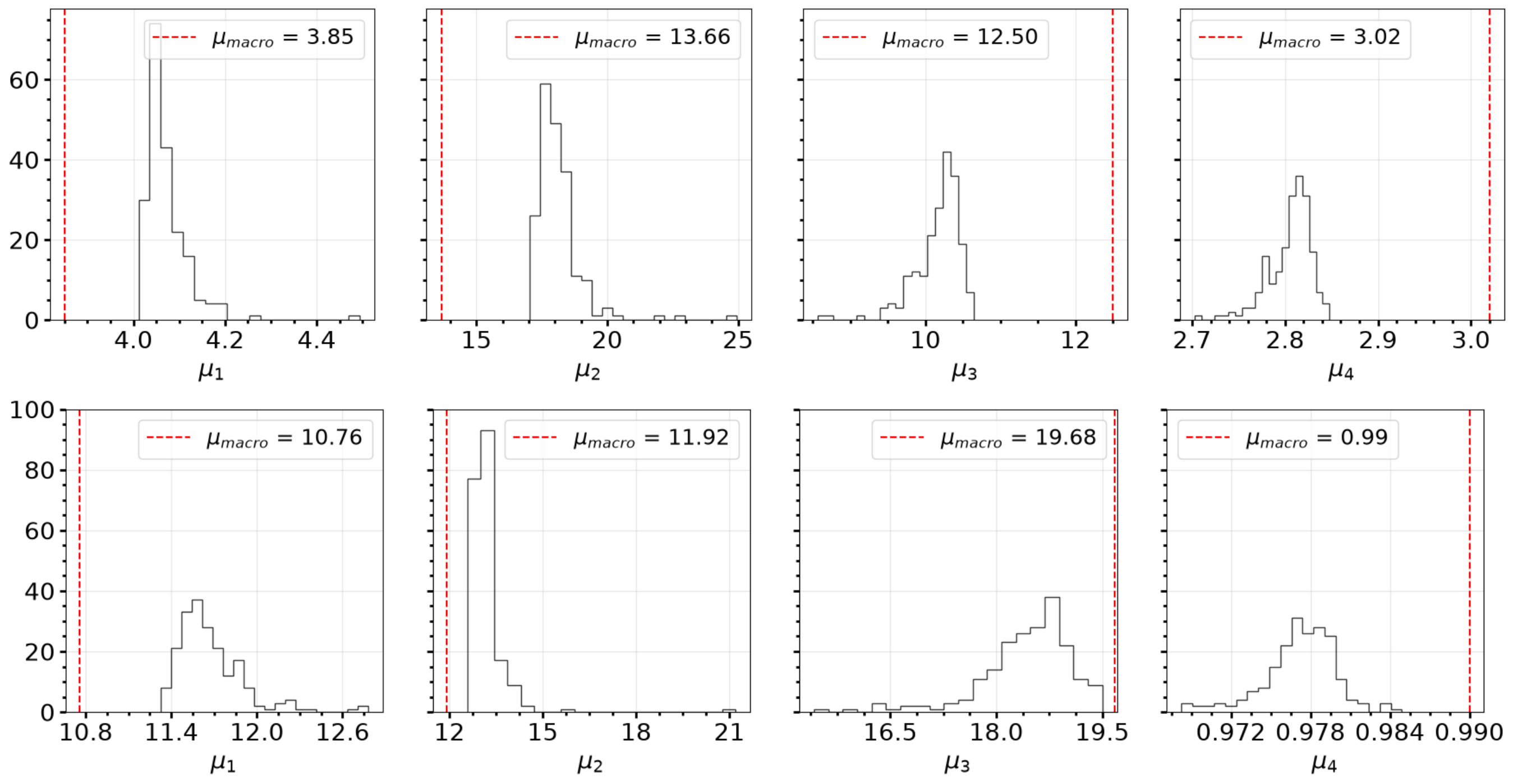}
    \caption{Absolute image magnifications due to 2NFW subhalos for PG 1115+080 (top) and RX J1131-1231 (below). The red line represents the no-subhalo case.}
    \label{mags}
\end{figure*}

Further, when incorporating the 2NFW subhalos, it is consistently observed that the absolute magnifications get substantially shifted (especially images 2 and 3) from the macro lens only magnifications (see Fig~\ref{mags}). An interesting feature is the narrower histogram for images 2 in both lenses (and Image 1 in PG1115+080), showing that magnifications do not vary much across realisations of the subhalo populations. Images 3 and 4 (and image 1 in RXJ1131-1231) on the other hand are broader indicating their higher sensitivity to subhalo distributions. Further, image parity is not observed to change, suggesting the NFW profile considered here is unlikely to cause this effect.
As with time delays, the effects on magnifications are also non-negligible and must be considered in lens modelling and cosmological parameter estimations. 

\subsection{Milli-images}
Apart from affecting the time delay of the macro-images, subhalos could also lead to multiple smaller milli-images being formed. In the context of observations, this would mean that we would observe multiple signals from these images in addition to the macro images, making this effect extremely important. 

We first consider the effect of a single NFW subhalo macro images 2 and 3 of the system PG 1115+080, varying its mass and distance from the images. 
Interestingly, no milli-images are formed for any mass-distance combination, with the macro images only getting perturbed, in the considered mass range. To confirm our results, milli-images were not found in any of the 200 realizations for the 2NFW subhalo population of both lenses, despite having over 10000 and 30000~subhalos, respectively. We verify our results using {\tt Lenstronomy} \citep{Birrer15, Birrer18, Birrer21} and can thus conclude that standard NFW subhalos are potentially unlikely to produce milli-images on their own.

Conversely, if milli-images are detected either in lensed EM or GW sources, then the density profiles of the subhalos are unlikely to be NFW. However, it is important to note that our conclusion applies to the standard NFW profile (with inner slope 1 and no core). Recent studies show that increasing the concentration \citep{Vujeva25} or including an inner core \citep{Williams24} in the NFW profile can produce milli-images under specific conditions. Other studies have shown that deviating from the NFW profile by increasing the inner slope can improve detectability and explain discrepancies in observations and simulations \citep{Kollmann26, Despali25}. The precision of lensed GWs provides a channel for detecting the tiny effects of the variations in the NFW model and accurately identifying the subhalo density profile.


\section{Conclusions}
\label{sec:conc}
In this study, we investigated the effect of dark matter subhalos on the time delays of images. We considered two types of lenses, a fold lens (PG 1115+080) and a cusp lens (RX J1131-1231), and replaced a fraction of the dark matter in the smooth component with subhalos. We defined two subhalo models: (i) point mass density with a pseudo-Jaffe spatial profile, and (ii) NFW density and spatial profile. 

Our results demonstrate that the choice of subhalo density profile has a significant impact on the magnitude of time-delay perturbations. Point mass subhalos produce large time delay perturbations of the order of several hours, and in the case of the cusp lens, even several days. The NFW subhalos, however, generate considerably small but substantial perturbations of a few hours, and approaching about a day for the cusp lens. The difference in the order of perturbations in the PMPJ and 2NFW models can be attributed to the shallower density of the NFW profile paired with the denser (towards the center and near the images) Pseudo-Jaffe spatial distribution. Additional examination of these models will be required to isolate the effects of the density profile and spatial distribution.   

We further investigated the ability of an NFW subhalo to produce milli images by placing individual subhalos close to macro images. No additional images were formed in our analysis. Moreover, simulations containing large populations of NFW subhalos likewise failed to generate milli-images. This finding suggests that standard NFW subhalos within the mass range explored here are unlikely to be responsible for the production of additional milli-images in strong-lensing systems. If milli-images are observed, they may hint at the presence of subhalos with modified density profiles, such as steeper inner slopes or enhanced central concentrations. Additionally, the order of time delay perturbations observed can conversely help constrain the density and spatial distributions of the dark matter substructure. We emphasize that the present analysis assumes a single density profile for all subhalos within a given realization. A more realistic description would require subhalo populations to be modelled as a mixture of different density profiles, which may involve luminous matter to be included too.

This study highlights the sensitivity of strong-lensing observables to the internal structure of dark matter subhalos and establishes a framework for future investigations of subhalo populations and their observational signatures. This work can be extended to different dark matter models, namely warm dark matter, self-interacting dark matter, etc. Future work will include modified NFW density profiles, such as profiles with steeper slopes and/or inner cores. The methodology can be readily extended to alternative dark matter scenarios, including warm dark matter and self-interacting dark matter models. The increasing precision of gravitational-wave detectors, together with forthcoming wide-field surveys such as the Vera C. Rubin Observatory's LSST, will provide opportunities to detect subhalo-induced perturbations in both gravitational-wave and electromagnetic lensing systems, opening new avenues for probing the nature of dark matter on sub-galactic scales.

\begin{acknowledgments}
AM would like to thank Sreekanth Harikumar for useful discussions.
\end{acknowledgments}

\section*{Data Availability}
The data underlying this article will be shared on reasonable request to the corresponding author.

\appendix*
\section{Spatial Distribution Effects}
\label{appendix}

\begin{figure*}
    \centering
    \includegraphics[scale = 0.15]{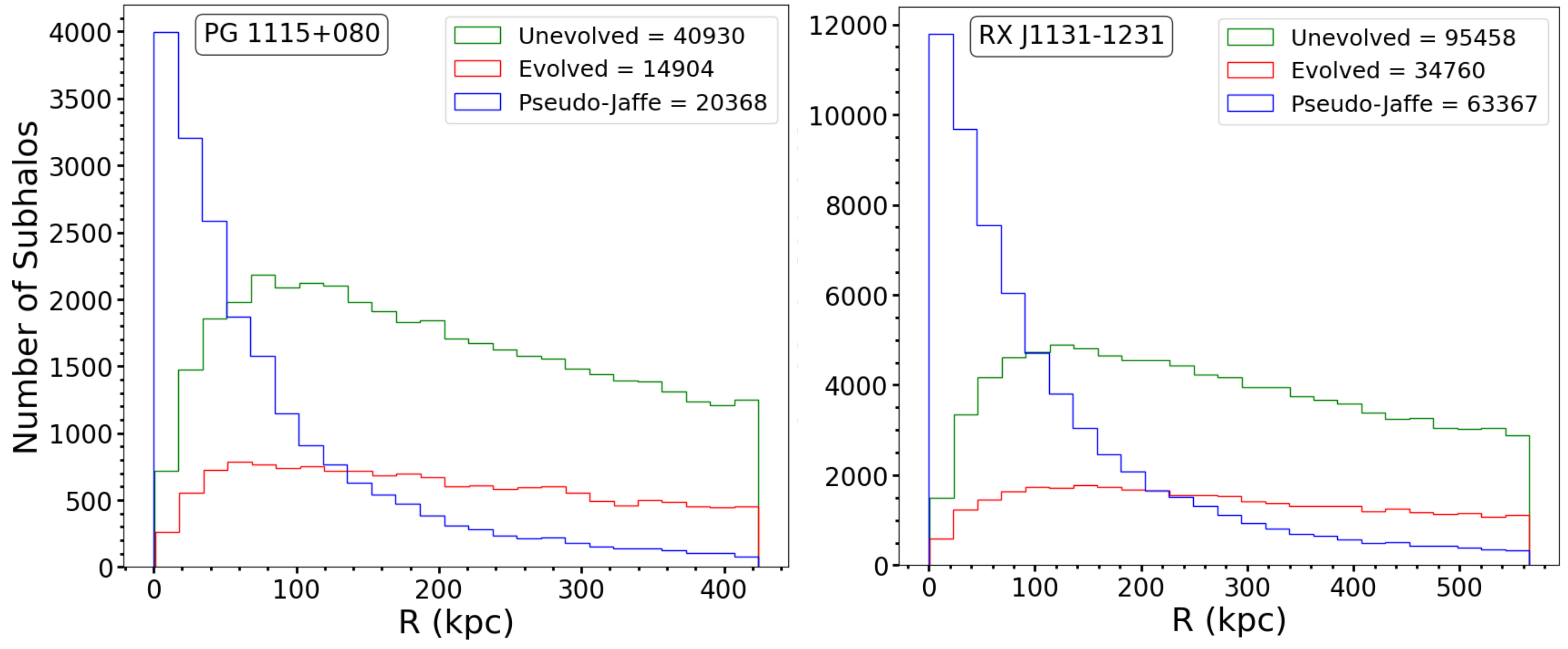}
    \caption{Comparison between Pseudo-Jaffe and NFW spatial distributions for PG 1115+080 (top) and RX J1131-1231 (bottom). For consistency, we set the same truncation radius for all three distributions.}
    \label{spdnfw}
\end{figure*}

\begin{figure*}
    \centering
    \includegraphics[width = \linewidth]{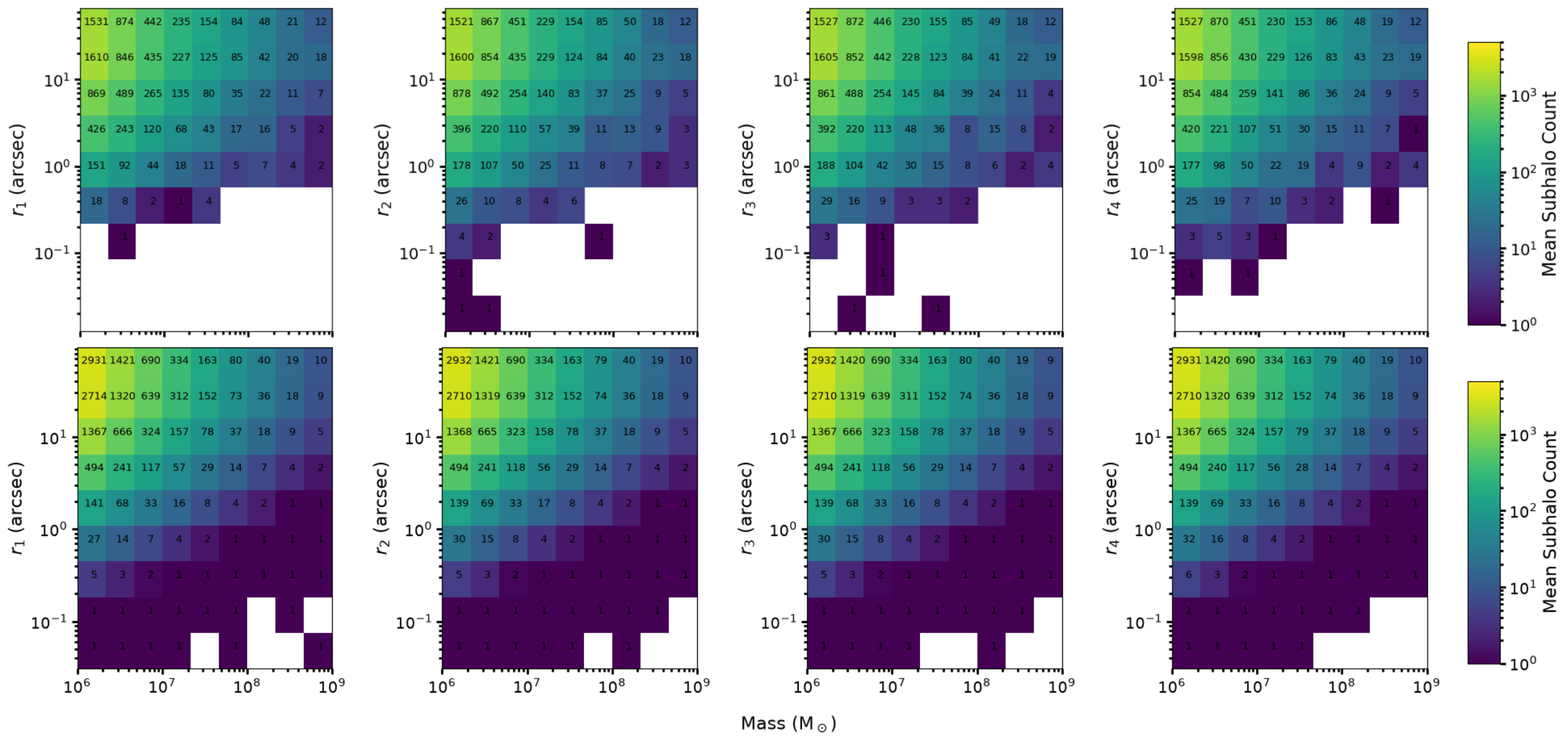}
    \caption{Comparison between PMPJ (top) and NFW (bottom) subhalo spatial distributions with their masses for the system PG 1115+080. The system RX J1131-1231 also results in the same conclusions and is hence omitted here.}
    \label{spdmcomp}
\end{figure*}

The Pseudo-Jaffe distribution evidently populates the majority of the subhalos near the lens centre and macro images, compared to the linear distribution of the 2NFW profile, explaining the higher perturbation in PMPJ model. This can be seen in Fig~\ref{spdnfw} which gives a comparison of the number of subhalos at different distances between the Pseudo-Jaffe spatial distribution and the NFW distribution (both evolved and unevolved). Furthermore, a comparison of the average number of subhalos of a certain mass at varying distances make it evident that PMPJ subhalos are denser near 1", the region where macro images exist, with some subhalos even closer than 0.1" to image 1 as visible in Fig~\ref{spdmcomp} \footnote{The results are an average over 200 realisations of PMPJ and 2NFW subhalos with their respective truncation radius and consequent properties. Subhalo counts are rounded off such that 0.2 is counted as 1.}. In addition to more subhalos near the images, there are more higher mass subhalos in this region, overall leading to greater perturbations. A simple extension of this work would be to spatial distribute NFW subhalos using Pseudo-Jaffe profile, to isolate the effects of just the spatial distribution.

\bibliography{citations}

@inbook{Dodelson03,
 author = {Dodelson, Scott},
 booktitle = {Modern Cosmology},
 isbn = {9780122191411},
 pages = {336--IV},
 publisher = {Elsevier},
 title = {{Analysis}},
 year = {2003}
}

@article{Planck15,
 author = {Ade, P. A. R. and others},
 issn = {1432-0746},
 journal = {Astronomy \& Astrophysics},
 month = {September},
 pages = {A13},
 publisher = {EDP Sciences},
 title = {{Planck2015} results: {XIII}. {Cosmological} parameters},
 url = {http://dx.doi.org/10.1051/0004-6361/201525830},
 volume = {594},
 year = {2016}
}

@article{planck20a,
 author = {Aghanim, N. and others},
 issn = {1432-0746},
 journal = {Astronomy \& Astrophysics},
 month = {September},
 pages = {A6},
 publisher = {EDP Sciences},
 title = {\textit{Planck} 2018 results: {VI}. {Cosmological} parameters},
 url = {http://dx.doi.org/10.1051/0004-6361/201833910},
 volume = {641},
 year = {2020}
}

@article{planck20b,
 author = {Akrami, Y. and others},
 issn = {1432-0746},
 journal = {Astronomy \& Astrophysics},
 month = {September},
 pages = {A7},
 publisher = {EDP Sciences},
 title = {\textit{Planck} 2018 results: {VII}. {Isotropy} and statistics of the {CMB}},
 url = {http://dx.doi.org/10.1051/0004-6361/201935201},
 volume = {641},
 year = {2020}
}

@article{Anderson14,
 author = {Anderson, Lauren and others},
 issn = {0035-8711},
 journal = {Monthly Notices of the Royal Astronomical Society},
 month = {April},
 number = {1},
 pages = {24--62},
 publisher = {Oxford University Press (OUP)},
 title = {{The} clustering of galaxies in the {SDSS-III} {Baryon} {Oscillation} {Spectroscopic} {Survey}: baryon acoustic oscillations in the {Data} {Releases} 10 and 11 {Galaxy} samples},
 url = {http://dx.doi.org/10.1093/mnras/stu523},
 volume = {441},
 year = {2014}
}

@article{DESI25,
 author = {Adame, A.G. and others},
 issn = {1475-7516},
 journal = {Journal of Cosmology and Astroparticle Physics},
 month = {February},
 number = {02},
 pages = {021},
 publisher = {IOP Publishing},
 title = {{DESI} 2024 {VI}: cosmological constraints from the measurements of baryon acoustic oscillations},
 url = {http://dx.doi.org/10.1088/1475-7516/2025/02/021},
 volume = {2025},
 year = {2025}
}

@ARTICLE{Schaye14,
  author={Schaye, Joop and others},
  journal={Monthly Notices of the Royal Astronomical Society}, 
  title={The EAGLE project: simulating the evolution and assembly of galaxies and their environments}, 
  year={2014},
  volume={446},
  number={1},
  pages={521-554},
  doi={10.1093/mnras/stu2058}}

@article{Vogel14,
  author={Vogelsberger, Mark and Genel, Shy and Springel, Volker and Torrey, Paul and Sijacki, Debora and Xu, Dandan and Snyder, Greg and Nelson, Dylan and Hernquist, Lars},
  journal={Monthly Notices of the Royal Astronomical Society}, 
  title={Introducing the Illustris Project: simulating the coevolution of dark and visible matter in the Universe}, 
  year={2014},
  volume={444},
  number={2},
  pages={1518-1547},
  doi={10.1093/mnras/stu1536}}

@article{Kravtsov10,
 author = {Kravtsov, Andrey},
 editor = {Hopp, Ulrich},
 issn = {1687-7977},
 journal = {Advances in Astronomy},
 month = {December},
 number = {1},
 publisher = {Wiley},
 title = {{Dark} {Matter} {Substructure} and {Dwarf} {Galactic} {Satellites}},
 url = {http://dx.doi.org/10.1155/2010/281913},
 volume = {2010},
 year = {2009}
}

@article{Kauffmann93,
 author = {Kauffmann, G. and White, S. D. M. and Guiderdoni, B.},
 issn = {1365-2966},
 journal = {Monthly Notices of the Royal Astronomical Society},
 month = {September},
 number = {1},
 pages = {201--218},
 publisher = {Oxford University Press (OUP)},
 title = {{The} formation and evolution of galaxies within merging dark matter haloes*},
 url = {http://dx.doi.org/10.1093/mnras/264.1.201},
 volume = {264},
 year = {1993}
}

@article{Klypin99,
 author = {Klypin, Anatoly and Kravtsov, Andrey V. and Valenzuela, Octavio and Prada, Francisco},
 issn = {1538-4357},
 journal = {The Astrophysical Journal},
 month = {September},
 number = {1},
 pages = {82--92},
 publisher = {American Astronomical Society},
 title = {{Where} {Are} the {Missing} {Galactic} {Satellites}?},
 url = {http://dx.doi.org/10.1086/307643},
 volume = {522},
 year = {1999}
}

@article{Moore99,
 author = {Moore, Ben and Ghigna, Sebastiano and Governato, Fabio and Lake, George and Quinn, Thomas and Stadel, Joachim and Tozzi, Paolo},
 issn = {0004-637X},
 journal = {The Astrophysical Journal},
 month = {October},
 number = {1},
 pages = {L19--L22},
 publisher = {American Astronomical Society},
 title = {{Dark} {Matter} {Substructure} within {Galactic} {Halos}},
 url = {http://dx.doi.org/10.1086/312287},
 volume = {524},
 year = {1999}
}

@article{Navarro96,
 author = {Navarro, Julio F. and Frenk, Carlos S. and White, Simon D. M.},
 issn = {1538-4357},
 journal = {The Astrophysical Journal},
 month = {May},
 pages = {563},
 publisher = {American Astronomical Society},
 title = {{The} {Structure} of {Cold} {Dark} {Matter} {Halos}},
 url = {http://dx.doi.org/10.1086/177173},
 volume = {462},
 year = {1996}
}

@article{Navarro10,
 author = {Navarro, Julio F. and Ludlow, Aaron and Springel, Volker and Wang, Jie and Vogelsberger, Mark and White, Simon D. M. and Jenkins, Adrian and Frenk, Carlos S. and Helmi, Amina},
 issn = {0035-8711},
 journal = {Monthly Notices of the Royal Astronomical Society},
 month = {December},
 number = {1},
 pages = {21--34},
 publisher = {Oxford University Press (OUP)},
 title = {{The} diversity and similarity of simulated cold dark matter haloes: {Diversity} and similarity of simulated {CDM} haloes},
 url = {http://dx.doi.org/10.1111/j.1365-2966.2009.15878.x},
 volume = {402},
 year = {2009}
}

@article{Stadel09,
 author = {Stadel, J. and Potter, D. and Moore, B. and Diemand, J. and Madau, P. and Zemp, M. and Kuhlen, M. and Quilis, V.},
 issn = {1745-3925},
 journal = {Monthly Notices of the Royal Astronomical Society: Letters},
 month = {September},
 number = {1},
 pages = {L21--L25},
 publisher = {Oxford University Press (OUP)},
 title = {{Quantifying} the heart of darkness with {GHALO} – a multibillion particle simulation of a galactic halo},
 url = {http://dx.doi.org/10.1111/j.1745-3933.2009.00699.x},
 volume = {398},
 year = {2009}
}

@article{Diemand08,
 author = {Diemand, J. and Kuhlen, M. and Madau, P. and Zemp, M. and Moore, B. and Potter, D. and Stadel, J.},
 issn = {1476-4687},
 journal = {Nature},
 month = {August},
 number = {7205},
 pages = {735--738},
 publisher = {Springer Science and Business Media LLC},
 title = {{Clumps} and streams in the local dark matter distribution},
 url = {http://dx.doi.org/10.1038/nature07153},
 volume = {454},
 year = {2008}
}

@article{Abazajian06,
 author = {Abazajian, Kevork},
 issn = {1550-2368},
 journal = {Phys. Rev. D},
 month = {March},
 number = {6},
 pages = {063513},
 publisher = {American Physical Society (APS)},
 title = {{Linear} cosmological structure limits on warm dark matter},
 url = {http://dx.doi.org/10.1103/physrevd.73.063513},
 volume = {73},
 year = {2006}
}

@article{Bode01,
 author = {Bode, Paul and Ostriker, Jeremiah P. and Turok, Neil},
 issn = {1538-4357},
 journal = {The Astrophysical Journal},
 month = {July},
 number = {1},
 pages = {93--107},
 publisher = {American Astronomical Society},
 title = {{Halo} {Formation} in {Warm} {Dark} {Matter} {Models}},
 url = {http://dx.doi.org/10.1086/321541},
 volume = {556},
 year = {2001}
}

@article{Spergel00,
 author = {Spergel, David N. and Steinhardt, Paul J.},
 issn = {1079-7114},
 journal = {Phys. Rev. Lett.},
 month = {April},
 number = {17},
 pages = {3760--3763},
 publisher = {American Physical Society (APS)},
 title = {{Observational} {Evidence} for {Self-Interacting} {Cold} {Dark} {Matter}},
 url = {http://dx.doi.org/10.1103/physrevlett.84.3760},
 volume = {84},
 year = {2000}
}

@article{Khoury15,
 author = {Khoury, Justin},
 issn = {1550-2368},
 journal = {Phys. Rev. D},
 month = {January},
 number = {2},
 pages = {024022},
 publisher = {American Physical Society (APS)},
 title = {{Alternative} to particle dark matter},
 url = {http://dx.doi.org/10.1103/physrevd.91.024022},
 volume = {91},
 year = {2015}
}

@article{Hsueh19,
 author = {Hsueh, J-W and Enzi, W and Vegetti, S and Auger, M W and Fassnacht, C D and Despali, G and Koopmans, L V E and McKean, J P},
 issn = {1365-2966},
 journal = {Monthly Notices of the Royal Astronomical Society},
 month = {November},
 number = {2},
 pages = {3047--3059},
 publisher = {Oxford University Press (OUP)},
 title = {{SHARP} – {VII}. {New} constraints on the dark matter free-streaming properties and substructure abundance from gravitationally lensed quasars},
 url = {http://dx.doi.org/10.1093/mnras/stz3177},
 volume = {492},
 year = {2019}
}

@article{Keeton01,
 title={Computational Methods for Gravitational Lensing},
 author={Charles R. Keeton},
 journal={arXiv: Astrophysics},
 year={2001},
 url={https://api.semanticscholar.org/CorpusID:7490470}
}

@article{Bullock01,
 author = {Bullock, James S. and Kravtsov, Andrey V. and Weinberg, David H.},
 issn = {1538-4357},
 journal = {The Astrophysical Journal},
 month = {February},
 number = {1},
 pages = {33--46},
 publisher = {American Astronomical Society},
 title = {{Hierarchical} {Galaxy} {Formation} and {Substructure} in the {Galaxy}’s {Stellar} {Halo}},
 url = {http://dx.doi.org/10.1086/318681},
 volume = {548},
 year = {2001}
}

@article{Navarro97,
doi = {10.1086/304888},
url = {https://doi.org/10.1086/304888},
year = {1997},
month = {dec},
publisher = {},
volume = {490},
number = {2},
pages = {493},
author = {Navarro, Julio F. and Frenk, Carlos S. and White, Simon D. M.},
title = {A Universal Density Profile from Hierarchical Clustering},
journal = {The Astrophysical Journal},
}

@article{Keeton03,
 author = {Keeton, Charles R.},
 issn = {1538-4357},
 journal = {The Astrophysical Journal},
 month = {February},
 number = {2},
 pages = {664--674},
 publisher = {American Astronomical Society},
 title = {{Analytic} {Cross} {Sections} for {Substructure} {Lensing}},
 url = {http://dx.doi.org/10.1086/345717},
 volume = {584},
 year = {2003}
}

@book{Schneider06,
 author = {Schneider, Peter and Kochanek, Christopher S. and Wambsganss, Joachim},
 isbn = {9783540303107},
 issn = {1861-7980},
 journal = {Saas-Fee Advanced Courses},
 publisher = {Springer Berlin Heidelberg},
 title = {{Gravitational} {Lensing}: {Strong}, {Weak} and {Micro}},
 year = {2006}
}

@article{Moustakas03,
 author = {Moustakas, L. A. and Metcalf, R. B.},
 issn = {1365-2966},
 journal = {Monthly Notices of the Royal Astronomical Society},
 month = {March},
 number = {3},
 pages = {607--615},
 publisher = {Oxford University Press (OUP)},
 title = {{Detecting} dark matter substructure spectroscopically in strong gravitational lenses},
 url = {http://dx.doi.org/10.1046/j.1365-8711.2003.06055.x},
 volume = {339},
 year = {2003}
}

@article{More09,
 author = {More, A. and McKean, J. P. and More, S. and Porcas, R. W. and Koopmans, L. V. E. and Garrett, M. A.},
 issn = {1365-2966},
 journal = {Monthly Notices of the Royal Astronomical Society},
 month = {March},
 number = {1},
 pages = {174--190},
 publisher = {Oxford University Press (OUP)},
 title = {{The} role of luminous substructure in the gravitational lens system {MG} 2016+112},
 url = {http://dx.doi.org/10.1111/j.1365-2966.2008.14342.x},
 volume = {394},
 year = {2009}
}

@article{Mao04,
 author = {Mao, Shude and Jing, Yipeng and Ostriker, Jeremiah P. and Weller, Jochen},
 issn = {1538-4357},
 journal = {The Astrophysical Journal},
 month = {March},
 number = {1},
 pages = {L5--L8},
 publisher = {American Astronomical Society},
 title = {{Anomalous} {Flux} {Ratios} in {Gravitational} {Lenses}: {For} or against {Cold} {Dark} {Matter}?},
 url = {http://dx.doi.org/10.1086/383413},
 volume = {604},
 year = {2004}
}

@article{Diemand07,
 author = {Diemand, Jurg and Kuhlen, Michael and Madau, Piero},
 issn = {1538-4357},
 journal = {The Astrophysical Journal},
 month = {March},
 number = {1},
 pages = {262--270},
 publisher = {American Astronomical Society},
 title = {{Dark} {Matter} {Substructure} and {Gamma}‐{Ray} {Annihilation} in the {Milky} {Way} {Halo}},
 url = {http://dx.doi.org/10.1086/510736},
 volume = {657},
 year = {2007}
}

@article{Hezaveh16,
 author = {Hezaveh, Yashar D. and Dalal, Neal and Marrone, Daniel P. and Mao, Yao-Yuan and Morningstar, Warren and Wen, Di and Blandford, Roger D. and Carlstrom, John E. and Fassnacht, Christopher D. and Holder, Gilbert P. and Kemball, Athol and Marshall, Philip J. and Murray, Norman and Levasseur, Laurence Perreault and Vieira, Joaquin D. and Wechsler, Risa H.},
 issn = {1538-4357},
 journal = {The Astrophysical Journal},
 month = {May},
 number = {1},
 pages = {37},
 publisher = {American Astronomical Society},
 title = {{DETECTION} {OF} {LENSING} {SUBSTRUCTURE} {USING} {ALMA} {OBSERVATIONS} {OF} {THE} {DUSTY} {GALAXY} {SDP}.81},
 url = {http://dx.doi.org/10.3847/0004-637X/823/1/37},
 volume = {823},
 year = {2016}
}

@article{Vegetti10,
 author = {Vegetti, S. and Koopmans, L. V. E. and Bolton, A. and Treu, T. and Gavazzi, R.},
 issn = {0035-8711},
 journal = {Monthly Notices of the Royal Astronomical Society},
 month = {October},
 number = {4},
 pages = {1969--1981},
 publisher = {Oxford University Press (OUP)},
 title = {{Detection} of a dark substructure through gravitational imaging: {Detection} of a dark substructure},
 url = {http://dx.doi.org/10.1111/j.1365-2966.2010.16865.x},
 volume = {408},
 year = {2010}
}

@article{Xu_15,
 author = {Xu, Dandan and Sluse, Dominique and Gao, Liang and Wang, Jie and Frenk, Carlos and Mao, Shude and Schneider, Peter and Springel, Volker},
 issn = {0035-8711},
 journal = {Monthly Notices of the Royal Astronomical Society},
 month = {January},
 number = {4},
 pages = {3189--3206},
 publisher = {Oxford University Press (OUP)},
 title = {{How} well can cold dark matter substructures account for the observed radio flux-ratio anomalies},
 url = {http://dx.doi.org/10.1093/mnras/stu2673},
 volume = {447},
 year = {2015}
}

@article{Gilman19a,
 author = {Gilman, Daniel and Birrer, Simon and Nierenberg, Anna and Treu, Tommaso and Du, Xiaolong and Benson, Andrew},
 issn = {1365-2966},
 journal = {Monthly Notices of the Royal Astronomical Society},
 month = {December},
 number = {4},
 pages = {6077--6101},
 publisher = {Oxford University Press (OUP)},
 title = {{Warm} dark matter chills out: constraints on the halo mass function and the free-streaming length of dark matter with eight quadruple-image strong gravitational lenses},
 url = {http://dx.doi.org/10.1093/mnras/stz3480},
 volume = {491},
 year = {2019}
}

@article{Gilman19b,
 author = {Gilman, Daniel and Du, Xiaolong and Benson, Andrew and Birrer, Simon and Nierenberg, Anna and Treu, Tommaso},
 issn = {1745-3933},
 journal = {Monthly Notices of the Royal Astronomical Society: Letters},
 month = {November},
 number = {1},
 pages = {L12--L16},
 publisher = {Oxford University Press (OUP)},
 title = {{Constraints} on the mass–concentration relation of cold dark matter haloes with 11 strong gravitational lenses},
 url = {http://dx.doi.org/10.1093/mnrasl/slz173},
 volume = {492},
 year = {2019}
}

@article{Gilman20,
 author = {Gilman, D. and Birrer, S. and Treu, T.},
 issn = {1432-0746},
 journal = {Astronomy \& Astrophysics},
 month = {October},
 pages = {A194},
 publisher = {EDP Sciences},
 title = {{TDCOSMO}: {III}. {Dark} matter substructure meets dark energy. {The} effects of (sub)halos on strong-lensing measurements of{H0}},
 url = {http://dx.doi.org/10.1051/0004-6361/202038829},
 volume = {642},
 year = {2020}
}

@article{Abe25,
 author = {Abe, Katsuya T. and Oguri, Masamune and Birrer, Simon and Khadka, Narayan and Marshall, Philip J. and Lemon, Cameron and More, Anupreeta and Collaboration, the LSST Dark Energy Science},
 issn = {2565-6120},
 journal = {The Open Journal of Astrophysics},
 month = {January},
 publisher = {Maynooth University},
 title = {{A} halo model approach for mock catalogs of time-variable strong gravitational lenses},
 url = {http://dx.doi.org/10.33232/001c.128482},
 volume = {8},
 year = {2025}
}

@article{Gannon25,
 title = {Dark matter substructure: A lensing perspective},
 author = {Gannon, Charles and Nierenberg, Anna and Benson, Andrew and Keeley, Ryan and Du, Xiaolong and Gilman, Daniel},
 journal = {Phys. Rev. D},
 volume = {112},
 issue = {2},
 pages = {023532},
 numpages = {17},
 year = {2025},
 month = {Jul},
 publisher = {American Physical Society},
 doi = {10.1103/kk2n-q4ps},
}

@article{Birrer25,
 author = {Birrer, Simon and Smith, Graham P. and Shajib, Anowar J. and Ryczanowski, Dan and Arendse, Nikki},
 issn = {1471-2962},
 journal = {Philosophical Transactions of the Royal Society A: Mathematical, Physical and Engineering Sciences},
 month = {May},
 number = {2295},
 publisher = {The Royal Society},
 title = {{Challenges} and opportunities for time-delay cosmography with multi-messenger gravitational lensing},
 url = {http://dx.doi.org/10.1098/rsta.2024.0130},
 volume = {383},
 year = {2025}
}

@article{Brehmer19,
 author = {Brehmer, Johann and Mishra-Sharma, Siddharth and Hermans, Joeri and Louppe, Gilles and Cranmer, Kyle},
 issn = {1538-4357},
 journal = {The Astrophysical Journal},
 month = {November},
 number = {1},
 pages = {49},
 publisher = {American Astronomical Society},
 title = {{Mining} for {Dark} {Matter} {Substructure}: {Inferring} {Subhalo} {Population} {Properties} from {Strong} {Lenses} with {Machine} {Learning}},
 url = {http://dx.doi.org/10.3847/1538-4357/ab4c41},
 volume = {886},
 year = {2019}
}

@article{Chiba02,
 author = {Chiba, Masashi},
 issn = {1538-4357},
 journal = {The Astrophysical Journal},
 month = {January},
 number = {1},
 pages = {17--23},
 publisher = {American Astronomical Society},
 title = {{Probing} {Dark} {Matter} {Substructure} in {Lens} {Galaxies}},
 url = {http://dx.doi.org/10.1086/324493},
 volume = {565},
 year = {2002}
}

@article{Dalal02,
 author = {Dalal, N. and Kochanek, C. S.},
 issn = {1538-4357},
 journal = {The Astrophysical Journal},
 month = {June},
 number = {1},
 pages = {25--33},
 publisher = {American Astronomical Society},
 title = {{Direct} {Detection} of {Cold} {Dark} {Matter} {Substructure}},
 url = {http://dx.doi.org/10.1086/340303},
 volume = {572},
 year = {2002}
}

@article{Mao98,
 author = {Mao, Shude and Schneider, Peter},
 issn = {1365-2966},
 journal = {Monthly Notices of the Royal Astronomical Society},
 month = {April},
 number = {3},
 pages = {587--594},
 publisher = {Oxford University Press (OUP)},
 title = {{Evidence} for substructure in lens galaxies?},
 url = {http://dx.doi.org/10.1046/j.1365-8711.1998.01319.x},
 volume = {295},
 year = {1998}
}

@article{Keeton09,
 author = {Keeton, Charles R. and Moustakas, Leonidas A.},
 issn = {1538-4357},
 journal = {The Astrophysical Journal},
 month = {June},
 number = {2},
 pages = {1720--1731},
 publisher = {American Astronomical Society},
 title = {{A} {NEW} {CHANNEL} {FOR} {DETECTING} {DARK} {MATTER} {SUBSTRUCTURE} {IN} {GALAXIES}: {GRAVITATIONAL} {LENS} {TIME} {DELAYS}},
 url = {http://dx.doi.org/10.1088/0004-637X/699/2/1720},
 volume = {699},
 year = {2009}
}

@article{Metcalf01,
 author = {Metcalf, R. Benton and Madau, Piero},
 issn = {1538-4357},
 journal = {The Astrophysical Journal},
 month = {December},
 number = {1},
 pages = {9--20},
 publisher = {American Astronomical Society},
 title = {{Compound} {Gravitational} {Lensing} as a {Probe} of {Dark} {Matter} {Substructure} within {Galaxy} {Halos}},
 url = {http://dx.doi.org/10.1086/323695},
 volume = {563},
 year = {2001}
}

@article{Liao17,
 author = {Liao, Kai and Fan, Xi-Long and Ding, Xuheng and Biesiada, Marek and Zhu, Zong-Hong},
 issn = {2041-1723},
 journal = {Nat. Commun.},
 month = {October},
 number = {1},
 publisher = {Springer Science and Business Media LLC},
 title = {{Precision} cosmology from future lensed gravitational wave and electromagnetic signals},
 url = {http://dx.doi.org/10.1038/s41467-017-01152-9},
 volume = {8},
 year = {2017}
}

@article{Minor21,
 author = {Minor, Quinn and Kaplinghat, Manoj and Chan, Tony H and Simon, Emily},
 issn = {1365-2966},
 journal = {Monthly Notices of the Royal Astronomical Society},
 month = {July},
 number = {1},
 pages = {1202--1215},
 publisher = {Oxford University Press (OUP)},
 title = {{Inferring} the concentration of dark matter subhaloes perturbing strongly lensed images},
 url = {http://dx.doi.org/10.1093/mnras/stab2209},
 volume = {507},
 year = {2021}
}

@article{Han16,
 author = {Han, Jiaxin and Cole, Shaun and Frenk, Carlos S. and Jing, Yipeng},
 issn = {1365-2966},
 journal = {Monthly Notices of the Royal Astronomical Society},
 month = {February},
 number = {2},
 pages = {1208--1223},
 publisher = {Oxford University Press (OUP)},
 title = {{A} unified model for the spatial and mass distribution of subhaloes},
 url = {http://dx.doi.org/10.1093/mnras/stv2900},
 volume = {457},
 year = {2016}
}

@article{Gilman19,
 author = {Gilman, Daniel and Birrer, Simon and Treu, Tommaso and Nierenberg, Anna and Benson, Andrew},
 issn = {1365-2966},
 journal = {Monthly Notices of the Royal Astronomical Society},
 month = {June},
 number = {4},
 pages = {5721--5738},
 publisher = {Oxford University Press (OUP)},
 title = {{Probing} dark matter structure down to 107 solar masses: flux ratio statistics in gravitational lenses with line-of-sight haloes},
 url = {http://dx.doi.org/10.1093/mnras/stz1593},
 volume = {487},
 year = {2019}
}

@article{Chen19,
 author = {Chen, Geoff C-F and Fassnacht, Christopher D and Suyu, Sherry H and Rusu, Cristian E and Chan, James H H and Wong, Kenneth C and Auger, Matthew W and Hilbert, Stefan and Bonvin, Vivien and Birrer, Simon and Millon, Martin and Koopmans, Léon V E and Lagattuta, David J and McKean, John P and Vegetti, Simona and Courbin, Frederic and Ding, Xuheng and Halkola, Aleksi and Jee, Inh and Shajib, Anowar J and Sluse, Dominique and Sonnenfeld, Alessandro and Treu, Tommaso},
 issn = {1365-2966},
 journal = {Monthly Notices of the Royal Astronomical Society},
 month = {September},
 number = {2},
 pages = {1743--1773},
 publisher = {Oxford University Press (OUP)},
 title = {{A} {SHARP} view of {H0LiCOW}: {H0} from three time-delay gravitational lens systems with adaptive optics imaging},
 url = {http://dx.doi.org/10.1093/mnras/stz2547},
 volume = {490},
 year = {2019}
}

@article{Birrer19,
 author = {Birrer, S and Treu, T and Rusu, C E and Bonvin, V and Fassnacht, C D and Chan, J H H and Agnello, A and Shajib, A J and Chen, G C-F and Auger, M and Courbin, F and Hilbert, S and Sluse, D and Suyu, S H and Wong, K C and Marshall, P and Lemaux, B C and Meylan, G},
 issn = {1365-2966},
 journal = {Monthly Notices of the Royal Astronomical Society},
 month = {January},
 number = {4},
 pages = {4726--4753},
 publisher = {Oxford University Press (OUP)},
 title = {{H0LiCOW} – {IX}. {Cosmographic} analysis of the doubly imaged quasar {SDSS} 1206+4332 and a new measurement of the {Hubble} constant},
 url = {http://dx.doi.org/10.1093/mnras/stz200},
 volume = {484},
 year = {2019}
}

@article{Wong19,
 author = {Wong, Kenneth C and Suyu, Sherry H and Chen, Geoff C-F and Rusu, Cristian E and Millon, Martin and Sluse, Dominique and Bonvin, Vivien and Fassnacht, Christopher D and Taubenberger, Stefan and Auger, Matthew W and Birrer, Simon and Chan, James H H and Courbin, Frederic and Hilbert, Stefan and Tihhonova, Olga and Treu, Tommaso and Agnello, Adriano and Ding, Xuheng and Jee, Inh and Komatsu, Eiichiro and Shajib, Anowar J and Sonnenfeld, Alessandro and Blandford, Roger D and Koopmans, Léon V E and Marshall, Philip J and Meylan, Georges},
 issn = {1365-2966},
 journal = {Monthly Notices of the Royal Astronomical Society},
 month = {September},
 number = {1},
 pages = {1420--1439},
 publisher = {Oxford University Press (OUP)},
 title = {{H0LiCOW} – {XIII}. {A} 2.4 per cent measurement of {H0} from lensed quasars: 5.3{$\sigma$} tension between early- and late-{Universe} probes},
 url = {http://dx.doi.org/10.1093/mnras/stz3094},
 volume = {498},
 year = {2019}
}

@ARTICLE{Shajib19,
 author = {{Shajib}, A.~J. and others},
 title = "{STRIDES: a 3.9 per cent measurement of the Hubble constant from the strong lens system DES J0408-5354}",
 journal = {Monthly Notices of the Royal Astronomical Society},
 year = 2020,
 month = jun,
 volume = {494},
 number = {4},
 pages = {6072-6102},
 doi = {10.1093/mnras/staa828},
archivePrefix = {arXiv},
 eprint = {1910.06306},
 primaryClass = {astro-ph.CO},
 adsurl = {https://ui.adsabs.harvard.edu/abs/2020MNRAS.494.6072S}
}

@article{Liao22,
 author = {Liao, Kai and Biesiada, Marek and Zhu, Zong-Hong},
 issn = {1741-3540},
 journal = {Chinese Physics Letters},
 month = {November},
 number = {11},
 pages = {119801},
 publisher = {IOP Publishing},
 title = {{Strongly} {Lensed} {Transient} {Sources}: {A} {Review}},
 url = {http://dx.doi.org/10.1088/0256-307X/39/11/119801},
 volume = {39},
 year = {2022}
}

@ARTICLE{Zentner03,
 author = {{Zentner}, Andrew R. and {Bullock}, James S.},
 title = "{Halo Substructure and the Power Spectrum}",
 journal = {The Astrophysical Journal},
 year = 2003,
 month = nov,
 volume = {598},
 number = {1},
 pages = {49-72},
 doi = {10.1086/378797},
archivePrefix = {arXiv},
 eprint = {astro-ph/0304292},
 primaryClass = {astro-ph},
 adsurl = {https://ui.adsabs.harvard.edu/abs/2003ApJ...598...49Z}
}

@article{Zentner05,
 author = {Zentner, Andrew R. and Berlind, Andreas A. and Bullock, James S. and Kravtsov, Andrey V. and Wechsler, Risa H.},
 issn = {1538-4357},
 journal = {The Astrophysical Journal},
 month = {May},
 number = {2},
 pages = {505--525},
 publisher = {American Astronomical Society},
 title = {{The} {Physics} of {Galaxy} {Clustering}. {I}. {A} {Model} for {Subhalo} {Populations}},
 url = {http://dx.doi.org/10.1086/428898},
 volume = {624},
 year = {2005}
}

@article{Vegetti12,
 author = {Vegetti, S. and Lagattuta, D. J. and McKean, J. P. and Auger, M. W. and Fassnacht, C. D. and Koopmans, L. V. E.},
 issn = {1476-4687},
 journal = {Nature},
 month = {January},
 number = {7381},
 pages = {341--343},
 publisher = {Springer Science and Business Media LLC},
 title = {{Gravitational} detection of a low-mass dark satellite galaxy at cosmological distance},
 url = {http://dx.doi.org/10.1038/nature10669},
 volume = {481},
 year = {2012}
}

@article{Sluse07,
 author = {Sluse, D. and Claeskens, J.-F. and Hutsemékers, D. and Surdej, J.},
 issn = {1432-0746},
 journal = {Astronomy \& Astrophysics},
 month = {March},
 number = {3},
 pages = {885--901},
 publisher = {EDP Sciences},
 title = {{Multi-wavelength} study of the gravitational lens system {RXS} {J1131-1231}: {III}. {Long} slit spectroscopy: micro-lensing probes the {QSO} structure},
 url = {http://dx.doi.org/10.1051/0004-6361:20066821},
 volume = {468},
 year = {2007}
}

@article{Sluse03,
 author = {Sluse, D. and Surdej, J. and Claeskens, J.-F. and Hutsemékers, D. and Jean, C. and Courbin, F. and Nakos, T. and Billeres, M. and Khmil, S. V.},
 issn = {1432-0746},
 journal = {Astronomy \& Astrophysics},
 month = {August},
 number = {2},
 pages = {L43--L46},
 publisher = {EDP Sciences},
 title = {{A} quadruply imaged quasar with an optical {Einstein} ring
candidate: 1{RXS} {J113155}.4–123155},
 url = {http://dx.doi.org/10.1051/0004-6361:20030904},
 volume = {406},
 year = {2003}
}

@article{Tonry98,
 author = {Tonry, John L.},
 issn = {0004-6256},
 journal = {The Astronomical Journal},
 month = {January},
 number = {1},
 pages = {1--5},
 publisher = {American Astronomical Society},
 title = {{Redshifts} of the {Gravitational} {Lenses} {B1422}+231 and {PG} 1115+080},
 url = {http://dx.doi.org/10.1086/300170},
 volume = {115},
 year = {1998}
}

@article{Weymann80,
 author = {Weymann, Ray J. and Latham, David and P. Angel, J. Roger and Green, Richard F. and Liebert, James W. and Turnshek, David A. and Turnshek, Diane E. and Tyson, J. Anthony},
 issn = {1476-4687},
 journal = {Nature},
 month = {June},
 number = {5767},
 pages = {641--643},
 publisher = {Springer Science and Business Media LLC},
 title = {{The} triple {QSO} {PG1115} + 08: another probable gravitational lens},
 url = {http://dx.doi.org/10.1038/285641a0},
 volume = {285},
 year = {1980}
}

@article{Yang21,
 author = {Yang, Lilan and Wu, Shichao and Liao, Kai and Ding, Xuheng and You, Zhiqiang and Cao, Zhoujian and Biesiada, Marek and Zhu, Zong-Hong},
 issn = {1365-2966},
 journal = {Monthly Notices of the Royal Astronomical Society},
 month = {November},
 number = {3},
 pages = {3772--3778},
 publisher = {Oxford University Press (OUP)},
 title = {{Event} rate predictions of strongly lensed gravitational waves with detector networks and more realistic templates},
 url = {http://dx.doi.org/10.1093/mnras/stab3298},
 volume = {509},
 year = {2021}
}

@ARTICLE{Keeton09a,
       author = {{Keeton}, Charles R.},
        title = "{Gravitational lensing with stochastic substructure: Effects of the clump mass function and spatial distribution}",
      journal = {arXiv e-prints},
         year = 2009,
        month = aug,
          eid = {arXiv:0908.3001},
        pages = {arXiv:0908.3001},
          doi = {10.48550/arXiv.0908.3001},
archivePrefix = {arXiv},
       eprint = {0908.3001},
 primaryClass = {astro-ph.CO},
       adsurl = {https://ui.adsabs.harvard.edu/abs/2009arXiv0908.3001K}
}

@article{Li18,
 author = {Li, Shun-Sheng and Mao, Shude and Zhao, Yuetong and Lu, Youjun},
 issn = {1365-2966},
 journal = {Monthly Notices of the Royal Astronomical Society},
 month = {February},
 number = {2},
 pages = {2220--2229},
 publisher = {Oxford University Press (OUP)},
 title = {{Gravitational} lensing of gravitational waves: a statistical perspective},
 url = {http://dx.doi.org/10.1093/mnras/sty411},
 volume = {476},
 year = {2018}
}

@article{Ding15,
 author = {Ding, Xuheng and Biesiada, Marek and Zhu, Zong-Hong},
 issn = {1475-7516},
 journal = {Journal of Cosmology and Astroparticle Physics},
 month = {December},
 number = {12},
 pages = {006--006},
 publisher = {IOP Publishing},
 title = {{Strongly} lensed gravitational waves from intrinsically faint double compact binaries—prediction for the {Einstein} {Telescope}},
 url = {http://dx.doi.org/10.1088/1475-7516/2015/12/006},
 volume = {2015},
 year = {2015}
}

@article{Pirkowska13,
 author = {Piórkowska, Aleksandra and Biesiada, Marek and Zhu, Zong-Hong},
 issn = {1475-7516},
 journal = {Journal of Cosmology and Astroparticle Physics},
 month = {October},
 number = {10},
 pages = {022--022},
 publisher = {IOP Publishing},
 title = {{Strong} gravitational lensing of gravitational waves in {Einstein} {Telescope}},
 url = {http://dx.doi.org/10.1088/1475-7516/2013/10/022},
 volume = {2013},
 year = {2013}
}

@article{Ng18,
 author = {Ng, Ken K. Y. and Wong, Kaze W. K. and Broadhurst, Tom and Li, Tjonnie G. F.},
 issn = {2470-0029},
 journal = {Phys. Rev. D},
 month = {January},
 number = {2},
 pages = {023012},
 publisher = {American Physical Society (APS)},
 title = {{Precise} {LIGO} lensing rate predictions for binary black holes},
 url = {http://dx.doi.org/10.1103/PhysRevD.97.023012},
 volume = {97},
 year = {2018}
}

@article{Rojo,
 title = {Gravitational lensing of waves: Novel methods for wave-optics phenomena},
 author = {Villarrubia-Rojo, Hector and Savastano, Stefano and Zumalac\'arregui, Miguel and Choi, Lyla and Goyal, Srashti and Dai, Liang and Tambalo, Giovanni},
 journal = {Phys. Rev. D},
 volume = {111},
 issue = {10},
 pages = {103539},
 numpages = {22},
 year = {2025},
 month = {May},
 publisher = {American Physical Society},
 doi = {10.1103/PhysRevD.111.103539},
 url = {https://link.aps.org/doi/10.1103/PhysRevD.111.103539}
}

@article{Sathya09,
 author = {Sathyaprakash, B. S. and Schutz, Bernard F.},
 issn = {1433-8351},
 journal = {Living Reviews in Relativity},
 month = {March},
 number = {1},
 publisher = {Springer Science and Business Media LLC},
 title = {{Physics}, {Astrophysics} and {Cosmology} with {Gravitational} {Waves}},
 url = {http://dx.doi.org/10.12942/lrr-2009-2},
 volume = {12},
 year = {2009}
}

@article{Regimbau12,
  author = {Regimbau, Tania and Dent, Thomas and Del Pozzo, Walter and Giampanis, Stefanos and Li, Tjonnie G. F. and Robinson, Craig and Van Den Broeck, Chris and Meacher, Duncan and Rodriguez, Carl and Sathyaprakash, B. S. and W{\'o}jcik, Katarzyna},
  issn = {1550-2368},
  journal = {Phys. Rev. D},
  number = {12},
  pages = {122001},
  publisher = {American Physical Society (APS)},
  title = {{Mock} data challenge for the {Einstein} {Gravitational-Wave} {Telescope}},
  url = {http://dx.doi.org/10.1103/PhysRevD.86.122001},
  volume = {86},
  year = {2012}
}

@article{Abbott17,
 author = {Abbott, B P and others},
issn = {1361-6382},
  journal = {Classical and Quantum Gravity},
  number = {4},
  pages = {044001},
  publisher = {IOP Publishing},
  title = {{Exploring the sensitivity of next generation gravitational wave detectors}},
  url = {http://dx.doi.org/10.1088/1361-6382/aa51f4},
  volume = {34},
  year = {2017}
}

@article{Oguri18,
 author = {Oguri, Masamune},
 issn = {1365-2966},
 journal = {Monthly Notices of the Royal Astronomical Society},
 month = {August},
 number = {3},
 pages = {3842--3855},
 publisher = {Oxford University Press (OUP)},
 title = {{Effect} of gravitational lensing on the distribution of gravitational waves from distant binary black hole mergers},
 url = {http://dx.doi.org/10.1093/mnras/sty2145},
 volume = {480},
 year = {2018}
}

@ARTICLE{Ishiyama21,
 author = {{Ishiyama}, Tomoaki and {Prada}, Francisco and {Klypin}, Anatoly A. and {Sinha}, Manodeep and {Metcalf}, R. Benton and {Jullo}, Eric and {Altieri}, Bruno and {Cora}, Sof{\'\i}a A. and {Croton}, Darren and {de la Torre}, Sylvain and et al.},
 title = "{The Uchuu simulations: Data Release 1 and dark matter halo concentrations}",
 journal = {Monthly Notices of the Royal Astronomical Society},
 year = 2021,
 month = sep,
 volume = {506},
 number = {3},
 pages = {4210-4231},
 doi = {10.1093/mnras/stab1755},
archivePrefix = {arXiv},
 eprint = {2007.14720},
 primaryClass = {astro-ph.CO},
 adsurl = {https://ui.adsabs.harvard.edu/abs/2021MNRAS.506.4210I}
}

@ARTICLE{Diemer18,
 author = {{Diemer}, Benedikt},
 title = "{COLOSSUS: A Python Toolkit for Cosmology, Large-scale Structure, and Dark Matter Halos}",
 journal = {The Astrophysical Journals},
 year = 2018,
 month = dec,
 volume = {239},
 number = {2},
 eid = {35},
 pages = {35},
 doi = {10.3847/1538-4365/aaee8c},
archivePrefix = {arXiv},
 eprint = {1712.04512},
 primaryClass = {astro-ph.CO},
 adsurl = {https://ui.adsabs.harvard.edu/abs/2018ApJS..239...35D}
}

@ARTICLE{Oguri10,
 author = {{Oguri}, Masamune},
 title = "{The Mass Distribution of SDSS J1004+4112 Revisited}",
 journal = {Publications of the Astronomical Society of Japan},
 year = 2010,
 month = aug,
 volume = {62},
 pages = {1017},
 doi = {10.1093/pasj/62.4.1017},
archivePrefix = {arXiv},
 eprint = {1005.3103},
 primaryClass = {astro-ph.CO},
 adsurl = {https://ui.adsabs.harvard.edu/abs/2010PASJ...62.1017O}
}

@misc{Vujeva25,
 archivePrefix = {arXiv},
 author = {Vujeva, Luka and Ezquiaga, Jose Mar\'{i}a and Gilman, Daniel and Goyal, Srashti and Zumalac\'{a}rregui, Miguel},
 copyright = {arXiv.org perpetual, non-exclusive license},
 eprint = {2510.14953},
 primaryClass = {astro-ph.CO},
 publisher = {arXiv},
 title = {{Dark} {Matter} {Subhalos} and {Higher} {Order} {Catastrophes} in {Gravitational} {Wave} {Lensing}},
 url = {https://arxiv.org/abs/2510.14953},
 year = {2025}
}

@article{Kollmann26,
 author = {Kollmann, Kassidy E and Nightingale, James W and Lisanti, Mariangela and Robertson, Andrew and Slone, Oren},
 issn = {1365-2966},
 journal = {Monthly Notices of the Royal Astronomical Society},
 month = {January},
 number = {4},
 pages = {stag066},
 publisher = {Oxford University Press (OUP)},
 title = {{Using} strong lensing to detect subhaloes with steep inner density profiles},
 url = {http://dx.doi.org/10.1093/mnras/stag066},
 volume = {546},
 year = {2026}
}

@ARTICLE{Williams24,
 author = {{Williams}, Liliya L.~R. and {Kelly}, Patrick L. and {Treu}, Tommaso and {Amruth}, Alfred and {Diego}, Jose M. and {Li}, Sung Kei and {Meena}, Ashish K. and {Zitrin}, Adi and {Broadhurst}, Thomas J. and {Filippenko}, Alexei V.},
 title = "{Flashlights: Properties of Highly Magnified Images Near Cluster Critical Curves in the Presence of Dark Matter Subhalos}",
 journal = {The Astrophysical Journal},
 year = 2024,
 month = feb,
 volume = {961},
 number = {2},
 eid = {200},
 pages = {200},
 doi = {10.3847/1538-4357/ad1660},
archivePrefix = {arXiv},
 eprint = {2304.06064},
 primaryClass = {astro-ph.CO},
 adsurl = {https://ui.adsabs.harvard.edu/abs/2024ApJ...961..200W}
}

@article{Birrer18,
 author = {Birrer, Simon and Amara, Adam},
 issn = {2212-6864},
 journal = {Physics of the Dark Universe},
 month = {December},
 pages = {189--201},
 publisher = {Elsevier BV},
 title = {lenstronomy: {Multi-purpose} gravitational lens modelling software package},
 url = {http://dx.doi.org/10.1016/j.dark.2018.11.002},
 volume = {22},
 year = {2018}
}

@article{Birrer21, doi = {10.21105/joss.03283}, url = {https://doi.org/10.21105/joss.03283}, year = {2021}, publisher = {The Open Journal}, volume = {6}, number = {62}, pages = {3283}, author = {Birrer, Simon and others}, title = {lenstronomy II: A gravitational lensing software ecosystem}, journal = {Journal of Open Source Software} }

@ARTICLE{Birrer15,
 author = {{Birrer}, Simon and {Amara}, Adam and {Refregier}, Alexandre},
 title = "{Gravitational Lens Modeling with Basis Sets}",
 journal = {The Astrophysical Journal},
 year = 2015,
 month = nov,
 volume = {813},
 number = {2},
 eid = {102},
 pages = {102},
 doi = {10.1088/0004-637X/813/2/102},
archivePrefix = {arXiv},
 eprint = {1504.07629},
 primaryClass = {astro-ph.CO},
 adsurl = {https://ui.adsabs.harvard.edu/abs/2015ApJ...813..102B}
}

@article{astropy:2013,
 Adsurl = {https://adsabs.harvard.edu/abs/2013A%26A...558A..33A},
 Archiveprefix = {arXiv},
 Author = {{Astropy Collaboration} and {Robitaille}, T.~P. and others},
 Doi = {10.1051/0004-6361/201322068},
 Eid = {A33},
 Eprint = {1307.6212},
 Journal = {Astronomy \& Astrophysics},
 Month = oct,
 Pages = {A33},
 Primaryclass = {astro-ph.IM},
 Title = {{Astropy: A community Python package for astronomy}},
 Volume = 558,
 Year = 2013}

@ARTICLE{astropy:2018,
 author = {{Astropy Collaboration} and {Price-Whelan}, A.~M. and
 others},
 title = "{The Astropy Project: Building an Open-science Project and Status of the v2.0 Core Package}",
 journal = {The Astrophysical Journal},
 year = 2018,
 month = sep,
 volume = {156},
 number = {3},
 eid = {123},
 pages = {123},
 doi = {10.3847/1538-3881/aabc4f},
 archivePrefix = {arXiv},
 eprint = {1801.02634},
 primaryClass = {astro-ph.IM},
 adsurl = {https://ui.adsabs.harvard.edu/abs/2018AJ....156..123A}
 }

@ARTICLE{astropy:2022,
 author = {{Astropy Collaboration} and {Price-Whelan}, Adrian M. and others},
 title = "{The Astropy Project: Sustaining and Growing a Community-oriented Open-source Project and the Latest Major Release (v5.0) of the Core Package}",
 journal = {The Astrophysical Journal},
 year = 2022,
 month = aug,
 volume = {935},
 number = {2},
 eid = {167},
 pages = {167},
 doi = {10.3847/1538-4357/ac7c74},
 archivePrefix = {arXiv},
 eprint = {2206.14220},
 primaryClass = {astro-ph.IM},
 adsurl = {https://ui.adsabs.harvard.edu/abs/2022ApJ...935..167A}
 }

@article{Despali25,
 author = {Despali, Giulia and Heinze, Felix M. and Fassnacht, Christopher D. and Vegetti, Simona and Spingola, Cristiana and Klessen, Ralf and Tajalli, Maryam},
 issn = {1432-0746},
 journal = {Astronomy \& Astrophysics},
 month = {July},
 pages = {A222},
 publisher = {EDP Sciences},
 title = {{Detecting} low-mass haloes with strong gravitational lensing: {II}. {Constraints} on the density profiles of two detected subhaloes},
 url = {http://dx.doi.org/10.1051/0004-6361/202451546},
 volume = {699},
 year = {2025}
}

@article{Springel08,
 author = {Springel, V. and Wang, J. and Vogelsberger, M. and Ludlow, A. and Jenkins, A. and Helmi, A. and Navarro, J. F. and Frenk, C. S. and White, S. D. M.},
 title = {The Aquarius Project: the subhaloes of galactic haloes},
 journal = {Monthly Notices of the Royal Astronomical Society},
 volume = {391},
 number = {4},
 pages = {1685-1711},
 year = {2008},
 month = {12},
 issn = {0035-8711},
 doi = {10.1111/j.1365-2966.2008.14066.x},
 url = {https://doi.org/10.1111/j.1365-2966.2008.14066.x},
 eprint = {https://academic.oup.com/mnras/article-pdf/391/4/1685/4881147/mnras0391-1685.pdf},
}

@ARTICLE{Navarro04,
 author = {{Navarro}, J.~F. and {Hayashi}, E. and {Power}, C. and {Jenkins}, A.~R. and {Frenk}, C.~S. and {White}, S.~D.~M. and {Springel}, V. and {Stadel}, J. and {Quinn}, T.~R.},
 title = "{The inner structure of {\ensuremath{\Lambda}}CDM haloes - III. Universality and asymptotic slopes}",
 journal = {Monthly Notices of the Royal Astronomical Society},
 year = 2004,
 month = apr,
 volume = {349},
 number = {3},
 pages = {1039-1051},
 doi = {10.1111/j.1365-2966.2004.07586.x},
archivePrefix = {arXiv},
 eprint = {astro-ph/0311231},
 primaryClass = {astro-ph},
 adsurl = {https://ui.adsabs.harvard.edu/abs/2004MNRAS.349.1039N}
}

@article{Gao04,
 author = {{Gao}, L. and {White}, S.~D.~M. and {Jenkins}, A. and {Stoehr}, F. and {Springel}, V.},
 title = "{The subhalo populations of {\ensuremath{\Lambda}}CDM dark haloes}",
 journal = {Monthly Notices of the Royal Astronomical Society},
 year = 2004,
 month = dec,
 volume = {355},
 number = {3},
 pages = {819-834},
 doi = {10.1111/j.1365-2966.2004.08360.x},
archivePrefix = {arXiv},
 eprint = {astro-ph/0404589},
 primaryClass = {astro-ph},
 adsurl = {https://ui.adsabs.harvard.edu/abs/2004MNRAS.355..819G}
}

@article{Ghigna00,
 author = {{Ghigna}, S. and {Moore}, B. and {Governato}, F. and {Lake}, G. and {Quinn}, T. and {Stadel}, J.},
 title = "{Density Profiles and Substructure of Dark Matter Halos: Converging Results at Ultra-High Numerical Resolution}",
 journal = {The Astrophysical Journal},
 year = 2000,
 month = dec,
 volume = {544},
 number = {2},
 pages = {616-628},
 doi = {10.1086/317221},
archivePrefix = {arXiv},
 eprint = {astro-ph/9910166},
 primaryClass = {astro-ph},
 adsurl = {https://ui.adsabs.harvard.edu/abs/2000ApJ...544..616G}
}

@article{Helmi02,
 title = {The phase-space structure of a dark-matter halo: Implications for dark-matter direct detection experiments},
 author = {Helmi, Amina and White, Simon D. M. and Springel, Volker},
 journal = {Phys. Rev. D},
 volume = {66},
 issue = {6},
 pages = {063502},
 numpages = {11},
 year = {2002},
 month = {Sep},
 publisher = {American Physical Society},
 doi = {10.1103/PhysRevD.66.063502},
 url = {https://link.aps.org/doi/10.1103/PhysRevD.66.063502}
}

\end{document}